%% file: sec_via_select.tex
\documentclass[journal]{IEEEtran}
\usepackage{blindtext}
\usepackage{etoolbox}
 
\makeatletter
\def\do#1{\patchcmd{#1}{\thepage}{\null}{}{\GenericWarning{}{Could not patch \string#1}}}
\docsvlist{\@oddhead,\@evenhead,\ps@headings,\ps@IEEEtitlepagestyle,\ps@IEEEpeerreviewcoverpagestyle}
\makeatother

\ifCLASSINFOpdf
\else
\fi
\usepackage{amsmath,dsfont,bbm,epsfig,amssymb,amsfonts}
\usepackage{amstext,verbatim,amsopn,cite,subfigure,multirow,multicol,lipsum,xfrac}
\usepackage{amsthm}
\usepackage[T1]{fontenc}
\usepackage{mathtools,amsthm}
\usepackage{perpage}
\usepackage{balance}
\usepackage{url}
\usepackage{amsfonts}
\usepackage{epsfig}
\usepackage{caption}
\usepackage{graphicx}
\usepackage{psfrag}	
\usepackage{etoolbox}
\usepackage{algorithmicx}
\usepackage[Algorithm,ruled]{algorithm}
\usepackage{algpseudocode}
\usepackage{pifont}
\usepackage[utf8]{inputenc}
\usepackage[T1]{fontenc}  
\usepackage[nolist]{acronym}
\MakePerPage{footnote}
\usepackage{paralist}
\usepackage{enumitem}
\usepackage{bbm}
\usepackage[process=auto]{pstool}
\usepackage{tikz}
\usetikzlibrary{shapes,arrows}

\usepackage{multicol}
\include{commands}
\begin{document}

\begin{acronym}
\acro{mimo}[MIMO]{Multiple-Input Multiple-Output}
\acro{mimome}[MIMOME]{Multiple-Input Multiple-Output Multiple-Eavesdropper}
\acro{csi}[CSI]{Channel State Information}
\acro{awgn}[AWGN]{Additive White Gaussian Noise}
\acro{iid}[i.i.d.]{independent and identically distributed}
\acro{ut}[UT]{User Terminal}
%\acro{bss}[BSs]{Base Stations}
\acro{bs}[BS]{Base Station}
\acro{tas}[TAS]{Transmit Antenna Selection}
\acro{lse}[LSE]{Least Squared Error}
\acro{rhs}[r.h.s.]{right hand side}
\acro{lhs}[l.h.s.]{left hand side}
\acro{wrt}[w.r.t.]{with respect to}
\acro{rs}[RS]{Replica Symmetry}
\acro{rsb}[RSB]{Replica Symmetry Breaking}
\acro{papr}[PAPR]{Peak-to-Average Power Ratio}
\acro{rzf}[RZF]{Regularized Zero Forcing}
\acro{snr}[SNR]{Signal-to-Noise Ratio}
\acro{rf}[RF]{Radio Frequency}
\acro{mf}[MF]{Match Filtering}
\acro{cdf}[CDF]{Cumulative Distribution Function}
\acro{tdd}[TDD]{Time Division Duplexing}
\end{acronym}

 % add this line after \begin{document} but before \titlepage

\title{Optimal Transmit Antenna Selection for Massive MIMO Wiretap Channels}

% Authors
\author{
\IEEEauthorblockN{
Saba Asaad,  \textit{Student Member, IEEE}, %\IEEEauthorrefmark{1}\IEEEauthorrefmark{2}, 
Ali Bereyhi,  \textit{Student Member, IEEE}, %\IEEEauthorrefmark{2}, 
Amir M. Rabiei, %\IEEEauthorrefmark{1}, 
Ralf R. M\"uller,  \textit{Senior Member, IEEE}, %\IEEEauthorrefmark{2}, 
Rafael F. Schaefer,  \textit{Senior Member, IEEE} %\IEEEauthorrefmark{3}
}
%\IEEEauthorblockA{
%\IEEEauthorrefmark{1}School of Electrical and Computer Engineering, University of Tehran\\
%\IEEEauthorrefmark{2}Institute for Digital Communications (IDC), Friedrich-Alexander Universit\"at Erlangen-N\"urnberg (FAU)\\
%\IEEEauthorrefmark{3}Information Theory and Applications Chair, Technische Universität Berlin (TUB)\\
%saba{\_}asaad@ut.ac.ir, ali.bereyhi@fau.de, ralf.r.mueller@fau.de, rafael.schaefer@tu-berlin.de, rabiei@ut.ac.ir
\thanks{This work has been presented in part at GLOBECOM 2017 \cite{asaad2017optimal}.}
\thanks{
%Saba Asaad is with the School of Electrical and Computer Engineering, University of Tehran and the Institute for Digital Communications, Friedrich-Alexander Universit\"at Erlangen-N\"urnberg (e-mail: saba{\_}asaad@ut.ac.ir).}
%\thanks{Ali Bereyhi is with the Institute for Digital Communications, Friedrich-Alexander Universit\"at Erlangen-N\"urnberg (e-mail: ali.bereyhi@fau.de).}
%\thanks{Amir Masoud Rabiei is with the School of Electrical and Computer Engineering, University of Tehran (e-mail: rabiei@ut.ac.ir).}
%\thanks{Ralf R. M\"uller is with the Institute for Digital Communications, Friedrich-Alexander Universit\"at Erlangen-N\"urnberg (e-mail: ralf.r.mueller@fau.de).}
%\thanks{Rafael F. Schaefer is with the Information Theory and Applications Chair, Technische Universit\"at Berlin (e-mail: rafael.schaefer@tu-berlin.de).}
%\thanks{This work was supported by the German Research Foundation, Deutsche Forschungsgemeinschaft (DFG), under Grant No. MU 3735/2-1.}
Saba Asaad and Amir Masoud Rabiei are with the School of Electrical and Computer Engineering (ECE), University of Tehran (UT), Tehran, Iran (e-mail: saba{\_}asaad@ut.ac.ir). Ali Bereyhi and Ralf R. M\"uller are with the Institute for Digital Communications (IDC), Friedrich-Alexander Universit\"at Erlangen-N\"urnberg (FAU), Erlangen, Germany (e-mail:ali.bereyhi@fau.de
and ralf.r.mueller@fau.de). Rafael
F. Schaefer is with the Information Theory and Applications Chair, Technische Universität Berlin (TUB), Berlin, Germany
(e-mail:rafael.schaefer@tu-berlin.de).}
\thanks{This work was supported by the German Research Foundation, Deutsche Forschungsgemeinschaft (DFG), under Grant No. MU 3735/2-1.}
}


\IEEEoverridecommandlockouts

\maketitle
  %Before starting of \begin{abstract}

\begin{abstract}
In this paper, we study the impacts~of~transmit~antenna selection on the secrecy performance of massive MIMO systems. We consider a wiretap setting in which a \textit{fixed} number of transmit antennas are selected and then confidential messages are transmitted over them to a multi-antenna legitimate receiver while being overheard by a multi-antenna eavesdropper. For this setup, we derive an accurate approximation of the instantaneous secrecy rate. Using this approximation, it is shown that in some wiretap settings under antenna selection the growth in the number of active antennas enhances the secrecy performance of the system up to some optimal number and degrades it when this optimal number is surpassed. This observation demonstrates that antenna selection in some massive MIMO settings not only reduces the RF-complexity, but also enhances the secrecy performance. We then consider various scenarios and derive the optimal number of active antennas analytically using our large-system approximation. Numerical investigations show an accurate match between simulations and the analytic results.
\end{abstract}
%ergodic secrecy rate, outage secrecy probability, optimal antenna selection. 
\IEEEpeerreviewmaketitle

%for providing reliable communication
% investigated the impact of employing multiple antennas in wiretap channel by studying the secrecy capacity of the \ac{mimome} channel.
%
%
%
%were proposed in the literature  to name just a few examples.
%To enhance the physical layer security, several lines of work have been considered in the literature investigating the impact of various concepts such as antenna selection \cite{yang2013transmit},  on the secrecy performance of \ac{mimo} systems.
%%
%
%
\section{Introduction}
Over the past few years, the popularity of smart phones, electronic tablets and video streaming as well as the sharp rise in the number of service providers has led to an explosive growth of data traffic in wireless networks. This increasing demand of capacity in mobile broadband communications poses challenges for designing the next generation of cellular networks (5G) in the near future \cite{agiwal2016next}. Given this backdrop, confidential and private transmission of data in the next generation of wireless networks is of paramount importance. In this respect, physical layer security for 5G wireless networks has gained significant attentions in recent years aiming for design of reliable and secure transmission schemes \cite{liu2017physical,zou2016survey}. Unlike the traditional approaches relying on cryptographic techniques \cite{massey1988introduction}, physical layer security provides secrecy by exploiting the inherent characteristics of wireless channels. Although cryptographic techniques employed in the upper layers of networks protect processed data securely, physical layer security is a potential solution through the communication phase\cite{yang2015safeguarding}.

The basic model for physical layer security is the wiretap channel in which transmitted messages to a legitimate receiver are being overheared by an eavesdropper. Wyner demonstrated that secrecy is obtained in this setting as long as the legitimate receiver communicates over a channel whose quality is better than the eavesdropper channel \cite{wyner1975wire}. Based on this framework, several techniques such as artificial noise generation \cite{deng2015security, wu2016secure} and cooperative jamming \cite{zheng2011optimal} were proposed for secrecy enhancement. The extension of Wyner's framework to \ac{mimo} settings has moreover shown a promising performance of such settings in the presence of eavesdroppers \cite{khisti2010secure,oggier2011secrecy,liu2009note}. In fact, in \ac{mimo} wiretap channels, also referred to as \ac{mimome} channels, the \ac{bs} can focus its main transmit beam to the legitimate terminals, and thus, reduce the information leakage to the eavesdroppers. This technique in massive \ac{mimo} settings \cite{marzetta2010noncooperative} asymptotically cancels out passive malicious terminals in the network making these settings robust against passive eavesdropping \cite{kapetanovic2015physical}. %Nevertheless, such a robustness vanishes when the eavesdropper is equipped with a large number of antennas \cite{zhu2014secure}. 

Despite promising characteristics of massive \ac{mimo} systems, they are known to pose high \ac{rf}-cost and complexity. In fact, employing a separate \ac{rf} chain per antenna in massive \ac{mimo} systems imposes a burden from the implementational point of view \cite{bjornson2015optimal}. This issue has introduced the antenna selection \cite{molisch2005capacity} along with other approaches such as spatial modulation \cite{di2014spatial} and hybrid analog-digital precoding schemes \cite{liang2014low, sohrabi2016hybrid} as prevalent strategies in massive \ac{mimo}. In antenna selection, only a subset of antennas is set~to~be active in each coherence time. This subset is in~general~selected with respect to some performance metric such as achievable transmission rate, outage probability or bit error rate \cite{molisch2005capacity}. The optimal approaches to antenna selection however deal with an exhaustive search which is not computationally feasible in practice. Alternatively, several suboptimal, but complexity efficient, methods have been proposed in the literature; see for example the approaches in \cite{gorokhov2003receive,gharavi2004fast,bereyhi2017asymptotics,bereyhi2018precoding}. The investigations have shown that these suboptimal approaches do not impose a significant loss on the performance for several \ac{mimo} settings \cite{molisch2004mimo,sanayei2004antenna,bereyhi2017asymptotics}. In the context~of~massive~\ac{mimo}~systems, recent studies have demonstrated that the large-system properties of these systems are maintained even via simple antenna selection algorithms \cite{li2014energy, asaad2017tas}. 
%\ac{mimo} settings, these suboptimal approaches have been shown to 
%
%antenna selection has been shown to provide advantages in terms of the overall \ac{rf}-cost, as well as the computational complexity, without significant loss in the performance \cite{molisch2004mimo,sanayei2004antenna}. 
%
%In conventional \ac{mimo} settings, antenna selection has been shown to provide advantages in terms of the overall \ac{rf}-cost, as well as the computational complexity, without significant loss in the performance \cite{molisch2004mimo,sanayei2004antenna}. 

In addition to implementational complexity reduction, antenna selection was also observed to be beneficial in \ac{mimo} systems with respect to some performance measures such as secrecy rate \cite{huang2015secure,yang2013transmit}, energy efficiency \cite{zhou2014energy} and effective rate \cite{al2017effective} in some special cases. For instance, it was shown in \cite{alves2011enhanced} that single \ac{tas}, i.e., only one trannsmit antenna being active, in a conventional \ac{mimo} setup can achieve high levels of security, especially when the total number of transmit antennas increases. The study was later extended in \cite{alves2012performance} to cases with multi-antenna eavesdroppers demonstrating that similar results hold also in these settings. In \cite{yang2013physical}, secure transmission in a general \ac{mimome} channel was investigated under single \ac{tas}. Such results were further extended in the literature for other \ac{mimome} settings. For example in \cite{wang2014secure}, secure transmission was studied for  Nakagami-$m$ fading channels under single \ac{tas}. The impacts of imperfect channel estimation and antenna correlation were also investigated in \cite{al2017secrecy}. The average secrecy rate and secrecy diversity analysis for a simple single \ac{tas} scheme was moreover studied in \cite{sadeque2013average,zhu2016secrecy}. In \cite{ferdinand2013effects}, \ac{tas} with outdated \ac{csi} was analyzed for scenarios with single-antenna receivers. The effect of single \ac{tas} at the \ac{bs} in the presence of randomly located eavesdroppers with a full-duplex receiver was moreover studied in \cite{chen2016secrecy}. In contrast to single \ac{tas}, the secrecy performance of \ac{mimome} channels under the multiple \ac{tas}, i.e., setting multiple transmit antennas to be active, has not yet been addressed in the literature. In fact under multiple \ac{tas}, the growth in the number of transmit antennas is beneficial to both the legitimate receiver and the eavesdropper, and therefore, its effect on the overall secrecy performance is not clear. This paper intends to study the impact of multiple \ac{tas} in massive \ac{mimome} settings.
\subsection*{Contributions and Organization}
We study the secrecy performance of a \ac{mimome} channel in which the \ac{bs} employs a computationally simple \ac{tas} algorithm to select a \textit{fixed} number of transmit antennas. For this setting, the distribution of the instantaneous secrecy rate in the large-system limit, i.e., when the number of transmit antennas grows large, is accurately approximated. This approximation is then utilized to investigate the secrecy performance in two different scenarios: Scenario (A) in which the eavesdropper's \ac{csi} is available at transmit side, and Scenario (B) in which the \ac{bs} does not know the eavesdropper's \ac{csi}. Our investigations demonstrate that in both scenarios, there exist cases in which the secrecy performance is optimized when the number active antennas are less than the total number of transmit antennas. In other words, the growth in the number of selected antennas in some cases enhances the secrecy performance up to an optimal value; however, it becomes destructive if the number of the active antennas surpasses this optimal value. Invoking our large-system results, we develop a framework to derive~analy- tically this optimal value. The consistency of our approach is then confirmed through numerical investigations.

The remaining parts of this manuscript is structured as~foll-ows: Section~\ref{sec:sys} describes the system model. In Section \ref{sec:result}, we conduct analyses for large dimensions. The impacts of~\ac{tas}~on the secrecy performance is investigated in Section \ref{sec:sec_enh} where we also give some numerical results and discussions. Finally, the concluding remarks are given in Section \ref{conclusion}. The proofs of the main theorems are moreover provided in the appendices.

\textit{Notations:} Throughout the paper, scalars, vectors and matrices are denoted by non-bold, bold lower case, and bold upper case letters, respectively. $\setC$ represents the complex plain. The Hermitian of $\mH$ is indicated with $\mH^{\her}$, and $\mI_N$ is the $N\times N$ identity matrix. The determinant of $\mH$ and Euclidean norm of $\bx$ are shown by $\abs{\mH}$ and $\norm{\bx}$, respectively. $\lfloor{x}\rceil$ refers to the integer with minimum Euclidean distance from $x$. The binary and natural logarithm are denoted by $\log\left(\cdot\right)$ and $\loge \left(\cdot\right)$, respectively, and $\mone_{\set{\cdot}}$ represents the indicator function. $\E\set{\cdot}$ is the mathematical expectation, and $\rmQ(x)$ and $\phi(x)$ denote the standard $\rmQ$-function and the zero-mean and unit-variance Gaussian distribution, respectively.

\section{Problem Formulation}
\label{sec:sys}
We consider a Gaussian \ac{mimome} wiretap setting in which the transmitter, the legitimate receiver and the eavesdropper are equipped with multiple antennas represented by $M$, $N_\rr$ and $N_\ee$, respectively. The main channel, from the transmitter to the legitimate receiver, and the eavesdropper~channel,~from the transmitter to the eavesdropper, are assumed to be statistically independent and experience quasi-static Rayleigh fading. The \ac{csi} of the both channels are considered to be available at the receiving terminals. The transmitter is moreover assumed to know the \ac{csi} of the main channel. In practice, the \ac{csi} is obtained at the respective terminals by performing channel estimation which depends on the duplexing mode of the system. Massive \ac{mimo} settings are usually considered to operate in the time division duplexing mode in which it is sufficient to estimate the channel only in the uplink training mode due to the channel reciprocity. More details on channel estimation in massive \ac{mimo} settings are found in \cite[Chapter 3]{marzetta2016fundamentals}. Based on the availability of the eavesdropper's \ac{csi} at the transmitter, we consider two different scenarios in this paper:
\begin{itemize}
\item[(A)] The eavesdropper's \ac{csi} is available at the transmitter.% knows the \ac{csi} of the eavesdropper channel.
\item[(B)] The transmitter does not know the eavesdropper's \ac{csi}.
\end{itemize}
%Next we consider the case in which the transmitter does not know the main channel \ac{csi}.
%In both channels, the channel coefficients are assumed to remain invariant during one block of transmission.
%This is a realistic assumption according to the slow fading characteristics of the channels. 
%denoted by the vector $\by_{N_\rr \times 1}$  which is written as
%
%
%\vspace*{-2mm}
\subsection{System Model}
\label{sec:tas}
The encoded message $\bx_{M \times 1}$ is transmitted over the main channel. In this case, the received signal $\by_{N_\rr \times 1}$ reads
\begin{align}
\by=\sqrt{\rho_\mm} \ \mH_\mm \bx + \bn_\mm \label{eq:sys-1}
\end{align}
where $\mH_\mm\hspace*{-1mm} \in\hspace*{-1mm} \setC^{N_\rr \times M}$ represents the main channel matrix,~$\rho_\mm$ denotes the average \ac{snr}~at~each~receive antenna and $\bn_{\mm}$ is zero-mean and unit-variance complex Gaussian noise, i.e., $\bn_\mm \sim \cnormal{\mI_{N_\rr}}$. Since the channel is assumed to be quasi-static Rayleigh fading, the coherence time is significantly larger than the transmission interval and entries of $\mH_\mm$ are modeled as \ac{iid} complex-valued Gaussian random variables with zero-mean and unit-variance.

At the eavesdropper, $\bx$ is overheard and the signal
\begin{align}
\bz=\sqrt{\rho_\ee} \ \mH_\ee \hspace{.2mm} \bx + \bn_\ee \label{eq:sys-2}
\end{align}
is received where $\mH_\ee \hspace*{-1mm} \in\hspace*{-1mm} \setC^{N_\ee \times M}$ is the eavesdropper channel matrix enclosing the fading coefficients between the transmit and eavesdropper's antennas. The entries if $\mH_\ee$ are modeled as \ac{iid} complex-valued Gaussian random variables with zero-mean and unit-variance, since the channel experiences quasi-static Rayleigh fading. $\bn_\ee \sim \cnormal{\mI_{N_\ee}}$ represents additive white Gaussian noise at the eavesdropper and  $\rho_\ee$ denotes the average \ac{snr} at each of the eavesdropper's antennas. While both the receiving terminals utilize all their available antennas, the transmitter employs the \ac{tas} protocol $\mas$ to select a subset of its antennas. The protocol is illustrated in the~following.
\textit{\ac{tas} Protocol:} Let $\bfh_{\ell, \mm} \hspace*{-1mm} \in\hspace*{-1mm} \setC^{N_\rr}$ represent the $\ell$-th column vector of $\mH_\mm$ and $L$ be the number of transmit antennas desired to be selected. Denote the index set of columns sorted with respect to their magnitudes by $\setW\coloneqq\left\lbrace w_1, \ldots, w_{M} \right\rbrace$ such that
\begin{align}
\norm{\bfh_{w_1, \mm}} \geq \norm{\bfh_{w_2, \mm}} \geq \cdots \geq \norm{\bfh_{w_{M}, \mm}}. \label{eq:sys-4}
\end{align}
The \ac{tas} protocol $\mas$ selects $L$ antennas which correspond to the first $L$ indices in $\setW$, i,e., $\setW_\mas\coloneqq \set{w_1, \ldots, w_{L}}$.

Corresponding to the \ac{tas} protocol, the effective~main~and eavesdropper channel, namely $\tmH_\mm$ and $\tmH_\ee$, are respectively constructed from $\mH_\mm$ and $\mH_\ee$ by collecting those column~vectors which correspond to the selected antennas. For instance, $\tmH_\mm$ is an $N_\mathrm{r} \times L$ matrix with columns $\bfh_{w_1, \mm}, \ldots, \bfh_{w_L, \mm}$. Note that although the \ac{tas} protocol $\mas$ selects the strongest antennas corresponding to the main channels, it performs as a random \ac{tas} protocol for the eavesdropper, since $\mH_\mm$ and $\mH_\ee$ are statistically independent.

\begin{remark}
In practice, $\setW_\mas$ in the \ac{tas} protocol can be determined either by employing a rate-limited feedback channel from the legitimate receiver to the \ac{bs} or by estimating the \ac{csi} at the \ac{bs}. One may note that in the former case, the rate-limited channel requires a low overhead. Moreover, in the latter case, the transmitter need not to acquire the complete \ac{csi}. In fact, as $\setW_\mas$ is determined via the ordering in \eqref{eq:sys-4}, the transmitter only needs to estimate the channel norms. This task can be done at the prior uplink stage simply by analog power estimators, and requires a significantly reduced time interval compared to the case of complete \ac{csi} estimation. This reduced interval furthermore allows for averaging over the coherence time which can improve the power estimation; see \cite{narasimhamurthy2009antenna} for more details.  %this measurements allow for averaging over a full coherence interval. Thus, the SNR of such a measurement is by a factor of T_c larger than the SNR of the data.
\end{remark}
\subsection{Achievable Secrecy Rate}
For the \ac{mimome} wiretap setting specified by \eqref{eq:sys-1} and \eqref{eq:sys-2}, the instantaneous achievable secrecy rate reads \cite{oggier2011secrecy}
\begin{align}
\mar_\rms =[\mar_\mm-\mar_\ee]^+\label{eq:sys-R_s}
%\\
%&=[\log\abs{\mI+\rho_\mm\mH_\mm \mQ \mH_\mm^\her}-   \log\abs{\mI+\rho_\ee\mH_\ee %\mQ \mH_\ee^\her}]^+ \label{eq:sys-5}
\end{align}
where $[x]^+=\max\set{0,x}$. In \eqref{eq:sys-R_s}, $\mar_\mm$ denotes the achievable rate over the main channel which reads
\begin{align}
\mar_\mm=\log\abs{\mI+\rho_\mm\mH_\mm \mQ \mH_\mm^\her} \label{eq:R_m}
\end{align}
and $\mar_\ee$ is the achievable rate over the eavesdropper channel which is given by
\begin{align}
\mar_\ee=\log\abs{\mI+\rho_\ee\mH_\ee \mQ \mH_\ee^\her} \label{eq:R_e}
\end{align} 
with $\mQ_{M \times M}$ being the power control matrix. For simplicity, we assume uniform power allocation over active antennas with unit average transmit power on each antenna. This means that %is set to be one
%,~i.e., %Considering the protocol $\mas$, $\mQ$ reads
\begin{align}
[\mQ]_{ww} = \left\{
        \begin{array}{ll}
            1 & w\in \setW_\mas \\
            0 & w \notin \setW_\mas
        \end{array}
    \right.
    .
\end{align}
Consequently, the instantaneous rates $\mar_\mm$ and $\mar_\ee$ reduce to% Considering the transmit power distribution,the maximum achievable rate of the main channel, $\mar_\mm$, under the \ac{tas} protocol reads in \eqref{eq:R_m} and \eqref{eq:R_e}
\begin{subequations}
\begin{align}
\mar_\mm &=\hspace*{.1mm}\log\abs{\mI+\rho_\mm\tmH_\mm^\her \tmH_{\mm}} \label{eq:R_mS} \\
\mar_\ee &=\log\hspace*{.1mm}\abs{\mI+\rho_\ee\tmH_\ee^\her \tmH_{\ee}}. \label{eq:R_eS}
\end{align}
\end{subequations}
Note that $\mar_\ee$ in \eqref{eq:R_eS} is determined under the~worst-case~scenario in which the eavesdropper knows the indices of antennas selected by the protocol $\mas$. Substituting into \eqref{eq:sys-R_s}, the maximum achievable instantaneous secrecy~rate~reads
\begin{align}
\mar_\rms\prnt{\mas}=\left[ \log\frac{\abs{\mI+{\rho_m}\tmH_\mm^\her \tmH_\mm}}{\abs{\mI+{\rho_\ee}\tmH_\ee^\her \tmH_\ee}}\right]^+  \label{eq:R_sS}
\end{align}
where the argument $\mas$ is written to indicate the dependency of the achievable secrecy rate on the \ac{tas} protocol. Note that when the eavesdropping terminal is not capable of obtaining the indices of the selected antennas, \eqref{eq:R_sS} bounds the achievable instantaneous secrecy rate from below. Since the channels experience fading, $\mar_\rms\prnt{\mas}$ is a random variable whose statistics define different secrecy performance metrics, e.g., the ergodic secrecy rate and secrecy outage probability. In the sequel, we evaluate the asymptotic distribution of $\mar_\rms\prnt{\mas}$.
%
%
%
%. For example, the expectation of $\mar_\rms\prnt{\mas}$ is defined to be the achievable ergodic secrecy rate which is an effective performance metric when the eavesdropper's \ac{csi} is known at the transmitter. Moreover, the \ac{cdf} of $\mar_\rms\prnt{\mas}$ determines the secrecy outage probability and is considered to quantify the secrecy performance of the system in Scenario~B. % and study the large-system secrecy performance in the presence of both active and passive eavesdroppers.
%
%Moreover, the argument $\mas$ in \eqref{eq:R_mS} and \eqref{eq:R_eS} indicate the dependency on the \ac{tas} protocol.
%
%which is an $N_\rr \times L$ matrix. Similarly, the maximum achievable rate of the eavesdropping channel $\mar_\ee$, under the \ac{tas} protocol is
%\begin{align}
%\mar_\ee=\log\abs{\mI+\rho_\ee\mH_\ee \mQ \mH_\ee^\her}=\log\abs{\mI+\frac{\rho_\ee}{L_t}\tmH_\ee^\her \tmH_\ee} 
%\end{align}
%where $\tmH_\ee$ is the effective eavesdropping channel which is an $N_\ee \times L$ matrix.
%In this case, the maximum achievable secrecy rate, when the \ac{tas} protocol is employed can be determined by substituting the effective channels $\tmH_\mm$ and $\tmH_\ee$ in \eqref{eq:sys-5} and distributing transmit power uniformly among the active transmit antennas.
% such that $\tr Q \leq 1$. In the case that the \ac{csi} is only available at the receiver, the achievable secrecy rate in \eqref{eq:sys-5} is maximized by setting equal power allocation amongst transmit antennas, i.e., $[\mQ]_{ii}=\dfrac{1}{N_t}$.
\section{Large-System Secrecy Performance}
\label{sec:result}
%Our main results give good approximations for $\mar_\rmE \prnt{\mas}$ and $\mar_\rmO\prnt{\mas;p_\rmO}$ in the large-system limits
%In this section, we investigate the asymptotics of the maximum achievable secrecy rate
%We investigate the secrecy performance of the system %\ac{mimome} wiretap channel under the \ac{tas} protocol~$\mas$ 
%by considering the following two cases.
%\begin{enumerate}[label= \Alph*]
%\item \label{itmA} The \ac{csi} of the eavesdropper channel is known at the transmitter. This case corresponds to scenarios in which the main channel is actively overheard.
%\item \label{itmB} The eavesdropper passively overhears the main channel. In this case, the \ac{csi} of the eavesdropper channel is not available at the transmit side.
%\end{enumerate}
%The eavesdropper is equipped with significantly fewer receive antennas compared to the number of~selected~antennas, i.e., $N_\ee \ll L$.
%The number of eavesdropper antennas grows large faster than the number of selected antennas, i.e., $N_\ee \gg L$.
%Case \ref{itmA} can be seen as a scenario in cellular networks with the eavesdropper being a regular user terminal. Moreover, Case \ref{itmB} describes a scenario in which the eavesdropper is a sophisticated terminal, such as portable stations.
%Cases \ref{itmA} and \ref{itmB} are usually referred to as active and passive eavesdropping, respectively and their 
%
%
The secrecy performances in Scenarios A and B are quantified via different metrics. In Scenario~A, since the \ac{bs} knows the eavesdropper's \ac{csi}, it transmits with rate $\mar_\rms\prnt{\mas}$ in each coherence time; thus, the secrecy performance is measured by the achievable ergodic secrecy rate. When the eavesdropper's \ac{csi} is not available at the \ac{bs}, the transmitter assumes the secrecy rate to be $\mar_\rmO$. In this case, the secure~transmission~is guaranteed as long as $\mar_\mm-\mar_\ee>\mar_\rmO$. Consequently,~in~Scenario~B, the secrecy performance is properly quantified by the secrecy outage capacity; see \cite{barros2006secrecy} for further discussions. 

%Therefore, the \ac{cdf} of $\mar_\rms(\mas)$ quantifies the secrecy performance. % of the system in this case. %probability of having information leakage when the transmitter has set the secrecy rate to $\mar_\rms$.   The secrecy outage probability, in the case of passive eavesdropping, is a fundamental metric in secrecy assessment.
%Using $\mar_\rms$ in \eqref{eq:sys-R_s}, different secrecy measures for the system is defined based on the eavesdropper's status. 

Based on above discussions, the performance of the setting in both Scenarios A and B is described by statistics~of~$\mar_\rms (\mas)$. We hence derive an accurate large-system approximation for the distribution of $\mar_\rms(\mas)$ in Theorem~\ref{thm:1}. Here by the large-system limit we mean $M\uparrow\infty$. To state Theorem~\ref{thm:1},~we~define the ``asymmetrically asymptotic regime of eavesdropping''.% in the following definition.
%
%
%by extending the results given in \cite{hochwald2004multiple,asaad2017tas} to our setup in Theorem~\ref{thm:1}. We then investigate the secrecy performance of the system in both scenarios using the large-system distribution of $\mar_\rms (\mas)$.
%
%
\begin{definition}[asymmetrically asymptotic regime of eavesdropping]
\label{def:rel}
The eavesdropper is said to overhear in the asymmetrically asymptotic regime of eavesdropping when the number of eavesdropper's antennas per active antenna,~defined~as~$\beta_\ee \coloneqq N_\ee / L$, reads either $\beta_\ee \ll 1$ or $\beta_\ee \gg 1$.
% close to zero or significantly larger than one. %In other words, we have $\beta_\ee \ll 1$ or $\beta_\ee \gg 1$.
\end{definition}
In Definition~\ref{def:rel}, $\beta_\ee \ll 1$ describes scenarios in which the eavesdropper is a regular mobile terminal with finite number of antennas. Moreover, $\beta_\ee \gg 1$ represents \ac{mimome} settings with sophisticated eavesdropping terminals such as portable stations in cellular networks. In the sequel, we assume that the understudy setting operates in the asymmetrically asymptotic regime of eavesdropping. However, our numerical investigations later depict that the results are valid even when~the~system does not operate in this   regime of eavesdropping.
%
%In the former case, the eavesdropper is equipped with significantly fewer receive antennas compared to the number of selected antennas, i.e., $N_\ee \ll L$ and the latter case corresponds to the scenario in which the number of eavesdropper antennas grows large faster than the number of selected antennas, i.e., $N_\ee \gg L$. 
%
%First we derive a good approximation of the maximum achievable secrecy rate in the large-system limit, i.e., $M\uparrow\infty$, then we study the impact of antenna selection on the secrecy performance of wireless system by considering different secrecy metrics such as ergodic secrecy rate, secrecy outage probability, $\epsilon$-outage secrecy capacity and asymptotic secrecy outage probability for assessing security for both cases of active eavesdropping and passive eavesdropping. 
%
% with the distribution of a the non-negative part of a Gaussian random variable.% and then in the next part optimal number of transit antennas in different scenarios is investigated.
%\begin{theorem}
%For the given setup, $\mar_\rms(\mas)$ is a random variable.
%%[Large-System Distribution of $\mar_\rms(\mas)$]
\begin{theorem}
\label{thm:1}
Consider the \ac{tas} protocol $\mas$, and let %$\eta_t$ and $\sigma_t^2$ 
\begin{subequations}
\begin{align}
\eta_t&=N_\mathrm{r}\left[ L  + M f_{N_\mathrm{r} +1}(u) \right] \label{eq:eta_t} \\
\sigma_t^2&=\left(uL-\eta_t \right)^2 \left(\frac{1}{L}-\frac{1}{M} \right) - \frac{\eta_t^2}{L}+ \Xi_t \label{eq:sigma_t}
\end{align}
\end{subequations}
for some non-negative real $u$ which satisfies
\begin{align}
\int_u^\infty f_{N_\mathrm{r}}(x) \mathrm{d} x= \frac{L}{M}, \label{eq:FIX-U}
\end{align}
$\Xi_t$ which is given by
\begin{align}
\Xi_t\coloneqq N_\mathrm{r} \left(N_\mathrm{r}+1\right) \left[ L+M f_{N_\mathrm{r}+1}(u)+M f_{N_\mathrm{r}+2}(u) \right],
\end{align}
and $f_{N_\mathrm{r}}(\cdot)$ which represents the chi-square probability density function with $2N_\mathrm{r}$ degrees of freedom and mean $N_\mathrm{r}$, i.e.,
 \begin{equation}
    f_{N_\mathrm{r}}(x)= \frac{1}{(N_\mathrm{r}-1)!}
    \begin{cases}
     x^{N_\mathrm{r}-1} e^{-x} , & \text{if}\ x \geq 0 \\
      0, & \text{if}\ x < 0
    \end{cases} \label{eq:DIST}
    .
  \end{equation}
Define the integers $U_\mm \coloneqq \min\set{L,N_\mathrm{r}}$, $V_\mm \coloneqq \max\set{L,N_\mathrm{r}}$, $U_\ee \coloneqq \min\set{L,N_\mathrm{e}}$ and $V_\ee \coloneqq \max\set{L,N_\mathrm{e}}$ and assume that the eavesdropper overhears in the asymmetrically asymptotic regime of eavesdropping. Then, as $M$ grows large, the distribution of the instantaneous secrecy rate $\mar_\rms(\mas)$ is effectively approximated by the distribution of $\mar_\mathrm{asy}(\mas)\coloneqq\left[ \mar^\star\right]^+$ where $\mar^\star$ is Gaussian with mean $\eta$ and variance $\sigma^2$ given by
\begin{subequations}
\begin{align}
\eta \coloneqq &U_\mm \log \left( \frac{K_t}{U_\mm}\right) \hspace*{-.7mm}-\hspace*{-.7mm} \dfrac{C_t \psi}{2 K_t^2} \rho_\mm \eta_t \hspace*{-.7mm} - \hspace*{-.7mm}U_\ee \log \left( 1+ \rho_\ee V_\ee \right), \label{eq:eta_final} \\
\sigma^2 \coloneqq &\left( \left[ \prnt{ 1-\dfrac{C_t }{ K_t^2} } {\frac{ U_\mm \rho_\mm \sigma_t}{K_t}} \right]^2  \right. \nonumber \\
&\left. + \mone_{\set{N_\ee > L}}\frac{U_\ee}{V_\ee} +\mone_{\set{N_\ee < L}} \frac{U_\ee V_\ee \rho^2_\ee}{\left(1+\rho_\ee V_\ee\right)^2  } \right) \psi^2 \label{eq:sigma_final}
\end{align}
\end{subequations}
for $K_t \coloneqq U_\mm + \rho_\mm \eta_t$ and $C_t \coloneqq \rho_\mm \eta_t U_\mm \prnt{U_\mm -1} / V_\mm$ and the constant $\psi = \log e = 1.4427.$
\end{theorem}
\begin{prf}
The proof follows the hardening property of the main and eavesdropper channel. In fact, the results of \cite{asaad2017tas} indicate that in the large-system limit, $\mar_\mm$ is approximately Gaussian with a vanishing variance. The eavesdropper~channel~is~moreover shown to harden in the asymmetrically~asymptotic~regi-me of eavesdropping following the discussions in \cite{hochwald2004multiple}. The detailed derivations are given in Appendix~\ref{app:1}.
%
%
%To evaluate the large-system distribution of $\mar_\rms(\mas)$, one needs to determine the asymptotic distribution of $\mar_\mm$ and $\mar_\ee$ given by \eqref{eq:R_m} and \eqref{eq:R_e} respectively. Using the results from \cite{asaad2017tas} and \cite{hochwald2004multiple}, the distribution of $\mar_\mm$ and $\mar_\ee$ in the large system limits for the both cases \ref{itmA} and \ref{itmB} can be approximated by Gaussian distributions. Noting the fact that the main and eavesdropper channels are independent, the random variable
%\begin{align}
%\mar^\star \coloneqq \mar_\mm-\mar_\ee, \label{eq:R_star}
%\end{align}
%in the large limit, can then be approximated with a Gaussian random variable whose variance and mean is determined in terms of the variances and means of $\mar_\mm$ and $\mar_\ee$. Finally by substituting in \eqref{eq:sys-R_s}, the proof is concluded. The detailed derivations are given in the appendix.
\end{prf}

From \eqref{eq:sigma_final}, one observes that the variance of the secrecy rate vanishes in the large-system limit. In fact, as $M$ grows large, $\eta_t$ increases, and hence, the first term in \eqref{eq:sigma_final} tends to zero. Moreover, in the asymmetrically asymptotic regime of eavesdropping, $U_\ee / V_\ee$ is significantly small and the two other terms are negligible. Consequently, in the large-system limit $\sigma$ converges to zero. This observation could be intuitively predicted, due to the fact that the both channels harden~asym-ptotically. The mean value $\eta$, however, does not necessarily increase as $M$ grows, since it is given as the difference of two terms which can both asymptotically grow large. The latter observation indicates that increasing the number of selected antennas for this setup does not necessarily improve the secrecy rate. We discuss this argument later in Section~\ref{sec:sec_enh}. At this point, we employ Theorem \ref{thm:1} to investigate the secrecy performance of the system in Scenarios~A and B. 
%Therefore, depending on the fluctuations of the main and eavesdropping channels over time, the ``ergodic secrecy~rate'' or ``secrecy outage probability''  is considered as the measure.
%The large-system approximation given in Theorem~\ref{thm:1}~enables us to discuss the asymptotics of different secrecy performance measures, such as ``ergodic secrecy~rate'' and ``secrecy outage probability'' which are being addressed in the sequel.
%. In the sequel, we investigate the asymptotic ergodic secrecy rate, as well as the asymptotic outage probability
\begin{remark}
Theorem~\ref{thm:1} gives a ``large-system approximation''. This means that for \textit{fixed} $L$, $N_\rr$ and $N_\ee$, $\mar_\mathrm{asy}(\mas)$ accurately approximates the statistics of the instantaneous secrecy rate when $M$ is large enough. Note that the theorem does not impose any constraint on the growth of $L$, $N_\rr$ and $N_\ee$, and the approximation is valid as long as the assumptions of the theorem are fulfilled. Nevertheless,~our~numerical~investigations show that even for $M=16$, which is not so large, this approximation is highly accurate.
\end{remark}
%the secrecy performance is characterized by the ergodic secrecy rate which is given by taking the expectation of $\mar_\rms\prnt{\mas}$. 
%, since the transmitter constructs the codewords by taking into account the given realization of the eavesdropper channel.
\subsection{Secrecy Performance in Scenario A}
\label{sec:active}
When the \ac{bs} knows the eavesdropper's \ac{csi}, the instantaneous secrecy rate is achievable in each transmission interval. Assuming that the symbols of the given codeword observe different realizations of the channel, the maximum average rate achieved by the transmitter is determined by the expectation of the instantaneous secrecy rate. This average rate is referred to as the achievable ergodic secrecy rate and is considered as an effective performance metric in this case. Using Theorem \ref{thm:1}, the achievable ergodic secrecy rate $\mar_\rmE \prnt{\mas}$ for our setup in the large-system limit is approximated as \begin{figure}[t]
\hspace*{-0.75cm}  
%%\centering
\resizebox{1.1\linewidth}{!}{
\pstool[width=.35\linewidth]{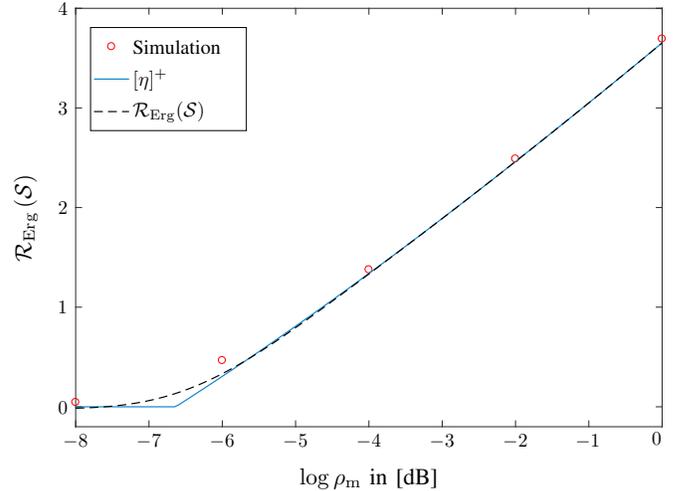}{
\psfrag{Rerg}[c][t][0.3]{$\mar_\rmE\prnt{\mas}$}
\psfrag{Lt}[c][c][0.3]{$\log \rho_\mm$ in [dB]}
\psfrag{Nt=1Nt=16-RR-NN}[l][l][0.26]{$\mar_\rmE(\mas)$}
\psfrag{Lt=8-Nt=16-EE-NN}[l][l][0.26]{$[\eta]^+$}
\psfrag{Nr=2-Ne=2-SS-NN}[l][l][0.26]{Simulation}
%\psfrag{Nt=256AAAB}[l][l][0.2]{$M=256$}

%y-axis
\psfrag{-8}[c][b][0.25]{$-8$}
\psfrag{-7}[c][b][0.25]{$-7$}
\psfrag{-6}[c][b][0.25]{$-6$}
\psfrag{-5}[c][b][0.25]{$-5$}
\psfrag{-4}[c][b][0.25]{$-4$}
\psfrag{-3}[c][b][0.25]{$-3$}
\psfrag{-2}[c][b][0.25]{$-2$}
\psfrag{-1}[c][b][0.25]{$-1$}
\psfrag{0}[c][b][0.25]{$0$}

%x-axis
\psfrag{0}[r][c][0.25]{$0$}
\psfrag{1}[r][c][0.25]{$1$}
\psfrag{2}[r][c][0.25]{$2$}
\psfrag{3}[r][c][0.25]{$3$}
\psfrag{4}[r][c][0.25]{$4$}
\psfrag{5}[r][c][0.25]{$5$}
}}
\caption{
%The ergodic secrecy rate in terms of the main channel's \ac{snr}. The curves have been plotted for $N_\rr=N_\ee=2$, $L=8$, $M=16$ and $\rho_\ee=-5$ dB. $\mar_\rmE(\mas)$ in \eqref{eq:asy} and the numerical simulations are sketched by a black line and red circles respectively. The blue line indicate $[\eta]^+$ for $\eta$ given in \eqref{eq:eta_final}. As it shows, the approximation tracks the simulations with high accuracy even for finite dimensions.\vspace*{-2mm}
The ergodic secrecy rate and its approximation versus the \ac{snr} of the legitimate receiver for $N_\rr=N_\ee=2$, $L=8$, $M=16$ and $\log \rho_\ee=-5$ dB.
%$\mar_\rmE(\mas)$ in \eqref{eq:asy} and the numerical simulations are sketched by a black line and red circles respectively. The blue line indicate $[\eta]^+$ for $\eta$ given in \eqref{eq:eta_final}. As it shows, the approximation tracks the simulations with high accuracy even for finite dimensions.\vspace*{-2mm}
}
\label{fig:1}
\vspace{-4.0mm}
\end{figure}
\begin{align}
\mar_\rmE \prnt{\mas} \approx \E \set{\mar_\asy\prnt{\mas}}&=\E \set{\left[\mar^\star\right]^+} \nonumber \\
%\int_{0}^{\infty} \phi\prnt{r;\eta,\sigma} r \dif r\\
&= \sigma \hspace*{.5mm} \phi\prnt{\xi}+ \eta \hspace*{.5mm}\rmQ\prnt{-\xi}. \label{eq:asy}
%\prnt{\mas} \coloneqq \E \set{\mar_\rms\prnt{\mas}}
\end{align}
where $\xi\coloneqq{\eta}/{\sigma}$. Using the inequality $\rmQ(x) < {\phi\prnt{x}}/{x}$ for $x>0$ and the fact that $\rmQ(-x)+\rmQ(x)=1$, we can bound the ergodic secrecy rate as
%\begin{align}
%\rmQ\prnt{-\xi} > 1-\frac{\sigma^2}{\eta} \phi\prnt{0;\eta,\sigma} 
%\end{align}
%main and eavesdropper channels fluctuate within the duration of a transmission interval, different transmissions observe different channel coefficients.
\begin{align}
\mar_\rmE\prnt{\mas} > \eta \label{eq:lower}
\end{align}
for $\xi>0$. By numerical investigations, it is seen that~the~lower bound is tight when $\xi$ is large enough. Fig.~\ref{fig:1} illustrates the accuracy of the approximations, as well as the tightness of the bound. The figure has been plotted for $L=8$~and~$M=16$ transmit antennas which is practically small. The \ac{snr} at the eavesdropping terminal is considered to be~$\log \rho_\ee=-5$~dB~and the receiving terminals have been assumed to have $N_\rr=N_\ee=2$ antennas. As the figure shows, the approximation is consistent with the simulations within a large range of \ac{snr}s. The lower bound in \eqref{eq:lower} moreover perfectly matches $\mar_\rmE(\mas)$ except for the interval of $\rho_\mm$ in which $\eta$ is close to zero. This observation is due to the fact that the variance in the large-system limit tends to zero rapidly, and thus, $\xi={\eta}/{\sigma}$ grows significantly large even for finite values of $\eta$. Consequently, one can write $\rmQ(-\xi)\approx 1-{ \phi(\xi)}/{\xi}$ and approximate the achievable ergodic rate with $\eta$ accurately. Although the approximation in Theorem~\ref{thm:1} is given for the large-system limit and asymmetrically asymptotic regime of eavesdropping, one observes that the result is accurately consistent with the simulations even for not so large dimensions and $\beta_\ee=1/8$.
\subsection{Secrecy Performance in Scenario B}
\label{sec:passive}
%\subsection{Secrecy Outage Probability}
%In slow fading scenarios~where~the channels do not fluctuate significantly within the whole transmission intervals, different transmissions observe the same realization of the channel coefficients.
%As the result, the achievable ergodic rate cannot be considered as an appropriate performance measure in this case.
In Scenario~B, the eavesdropper's \ac{csi} is not known at the \ac{bs}. This means that for a given realization of the~channels,~the instantaneous secrecy rate in \eqref{eq:sys-R_s} cannot be achieved. This is due to the fact that the transmitter achieves the secrecy rate in \eqref{eq:sys-R_s} by constructing its codewords based on the leakage rate achievable over the eavesdropper channel. For this scenario, the $\epsilon$-outage secrecy rate is known to be the proper metric quantifying the secrecy performance. Considering a given rate $\mar_\rmO\geq 0$ the secrecy outage probability $\Pout_\out \prnt{\mar_\rmO}$ is \cite{barros2006secrecy}
\begin{align}
\Pout_\out \prnt{\mar_\rmO} = \Pr\set{\mar_s \prnt{\mas} < \mar_\rmO }.  \label{eq:secrecy_outage}
\end{align}
Consequently, the $\epsilon$-outage achievable secrecy rate $\mar_\out \prnt{\epsilon}$ is defined as the maximum possible rate for which $\Pout_\out \prnt{\mar_\rmO} \leq \epsilon$. The intuition behind defining the $\epsilon$-outage secrecy rate as the performance metric can be stated as the following: Since the \ac{bs} does not know the \ac{csi} of the eavesdropper channel, it assumes that the achievable secrecy rate is at least $\mar_\rmO$ in all transmission intervals. Noting that the \ac{csi} of the main channel is known at the \ac{bs}, the setting of the secrecy rate implicitly imposes this assumption on the quality of the eavesdropper channel that $\mar_\ee < \mar_\mm-\mar_\rmO$ in which the term $\mar_\mm-\mar_\rmO$ is known by the transmitter. Consequently, the secrecy outage probability in \eqref{eq:secrecy_outage} determines the probability of the eavesdropper having better channel quality than the assumed term $\mar_\mm-\mar_\rmO$, or equivalently, the fraction of intervals in which the eavesdropper can decode transmit code-words at least partially. As a result, $\mar_\out\prnt{\epsilon}$ determines the maximum achievable secrecy rate for which one can guarantee that the fraction of transmission intervals with information be-ing leaked to the eavesdropper is less than $\epsilon$.

From Theorem~\ref{thm:1}, the outage probability is approximated~as
\begin{align}
\Pout_\out \prnt{\mar_\rmO} \approx &\Pr\set{\mar_\asy \prnt{\mas} \leq \mar_\rmO } =\Pr\set{\mar^\star \leq \mar_\rmO } \nonumber \\
&= 1- \rmQ\prnt{\frac{\mar_\rmO - \eta}{\sigma}}. \label{eq:out}
\end{align}
Consequently, the $\epsilon$-outage secrecy rate is given by
\begin{align}
\mar_{\out} \prnt{\epsilon} = \sigma \rmQ^{-1} \prnt{1-\epsilon } +\eta \label{eq:R_out-Final}
\end{align}
with $\rmQ^{-1}\prnt{\cdot}$ being the inverse of the $\rmQ$-function with respect to composition. 
%Using the notation of secrecy outage probability
Moreover, the probability of non-zero secrecy rate $\Pout_{\nzs}$, defined as $\Pout_{\nzs} \coloneqq \Pr\set{\mar_\rms \prnt{\mas} > 0 }$, in the large-system limit is approximated as $\Pout_{\nzs} \approx 1- \rmQ\prnt{{\eta}/{\sigma}}$.

\begin{figure}[t]
\hspace*{-0.75cm}  
%%\centering
\resizebox{1.1\linewidth}{!}{
\pstool[width=.35\linewidth]{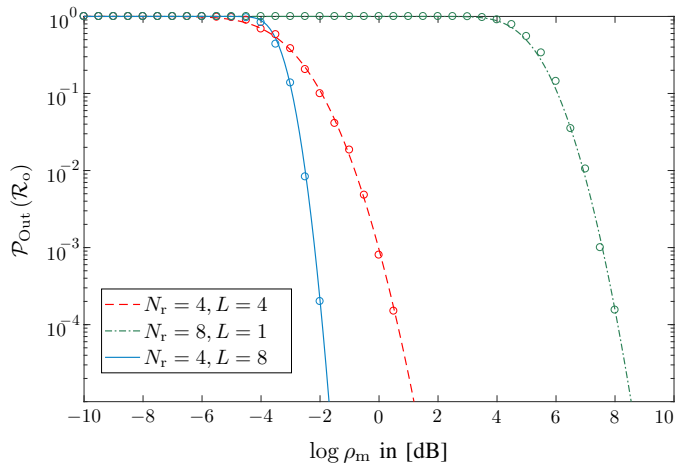}{
\psfrag{snr}[c][t][0.3]{$\log \rho_\mm$ in [dB]}
\psfrag{Outage Probability}[c][t][0.28]{$\Pout_\out \prnt{\mar_\rmO}$}
\psfrag{Nr=4, Ne=8, Lt=4AAAAA}[l][l][0.27]{$N_\rr=4, L=4$}
\psfrag{Nr=8, Ne=8, Lt=1AAAAA}[l][l][0.27]{$N_\rr=8, L=1$}
\psfrag{Nr=4, Ne=8, Lt=8AAAAA}[l][l][0.27]{$N_\rr=4, L=8$}
%\psfrag{Nt=256AAAB}[l][l][0.2]{$M=256$}
%
%y-axis
\psfrag{-10}[c][b][0.25]{$-10$}
\psfrag{-8}[c][b][0.25]{$-8$}
\psfrag{-7}[c][b][0.25]{$-7$}
\psfrag{-6}[c][b][0.25]{$-6$}
\psfrag{-5}[c][b][0.25]{$-5$}
\psfrag{-4}[r][t][0.2]{$-4\hspace*{-1mm}$}
\psfrag{-3}[r][t][0.2]{$-3\hspace*{-1mm}$}
\psfrag{-2}[r][t][0.2]{$-2\hspace*{-1mm}$}
\psfrag{-1}[r][t][0.2]{$-1\hspace*{-1mm}$}
\psfrag{0}[r][t][0.2]{$0\hspace*{-.2mm}$}
\psfrag{2}[c][b][0.25]{$2$}
\psfrag{4}[c][b][0.25]{$4$}
\psfrag{6}[c][b][0.25]{$6$}
\psfrag{8}[c][b][0.25]{$8$}
\psfrag{10}[r][b][0.25]{$10$}
\psfrag{-4x}[c][b][0.25]{$-4$}
\psfrag{-3x}[c][b][0.25]{$-3$}
\psfrag{-2x}[c][b][0.25]{$-2$}
\psfrag{-1x}[c][b][0.25]{$-1$}
\psfrag{0x}[c][b][0.25]{$0$}
%x-axis
}}

\vspace*{-4mm}\caption{The secrecy outage probability at $\mar_\rmO=5$ versus $\rho_\mm$ when $M=128$, $N_\ee=8$ and $\log \rho_\ee=-10$ dB. The solid lines and the circles, following each other closely, are the approximation and the simulated values, respectively.}
%\vspace{-4.0mm}
\label{fig:out}
\end{figure}

Fig.~\ref{fig:out} shows the secrecy outage probability as a function of $\rho_\mm$ for $\mar_\rmO=5$ considering various values of $N_\rr$ and $L$. Here, $N_\ee=8$ and $\log \rho_\ee=-10$ dB and the \ac{bs}~is~considered~to~be~equ-ipped with $M=128$ antennas. As it is seen, the large-system approximation consistently tracks the numerical result for a large range of \ac{snr}s. Although Theorem~\ref{thm:1} approximates the distribution of the instantaneous secrecy rate in the asymmetrically asymptotic regime of eavesdropping, one can see that the results closely match the simulations even for $\beta_\ee=1$.
%
%The closed-form characterization of the asymptotic secrecy performance given by Theorem~\ref{thm:1} enables us to investigate the impact of \ac{tas} on the performance of the system. We show that for the considered setup, the secrecy performance of the system in some cases is optimized for $L < M$. For these cases, we employ our analytic results to optimally select the number of active antennas. Moreover, we discuss the set of constraints in which the optimal secrecy performance is given by choosing a subset of transmit antennas.
%
%
%
%
%Considering either the ergodic secrecy rate or the secrecy outage probability, the secrecy performance of the system in the large limit is mainly specified by $\eta$. In contrast to $\sigma$ which tends to zero in the asymptotic regime, the factor $\eta$, for a given number of receive and eavesdropper antennas, can either grow, vanish or tend to some constant in the large-system limit depending on the number of selected antennas. 
%
\section{Secrecy Enhancement via \ac{tas}}
\label{sec:sec_enh}
\begin{figure}[t]
\hspace*{-0.75cm}  
%%\centering
\resizebox{1.1\linewidth}{!}{
\pstool[width=.35\linewidth]{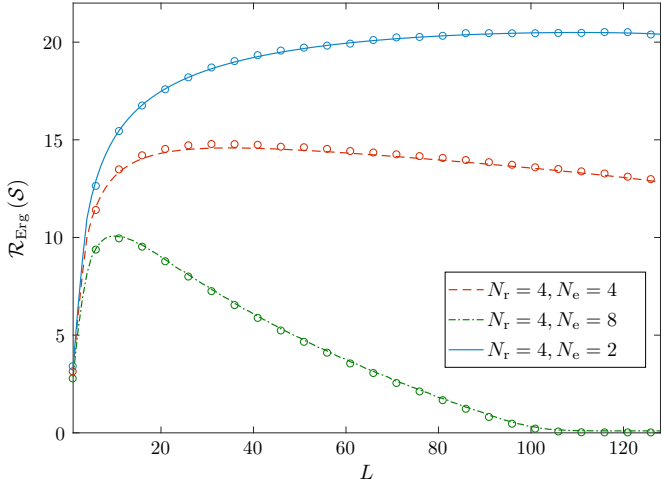}{
\psfrag{ErgCapacity}[c][t][0.27]{$\mar_\rmE\prnt{\mas}$}
\psfrag{Lt}[c][c][0.27]{$L$}
\psfrag{Nr=4AAABBB, Ne=2AAAA}[l][l][0.26]{$N_\rr=4, N_\ee=2$}
\psfrag{Nr=4AAABBB, Ne=4AAAA}[l][l][0.26]{$N_\rr=4, N_\ee=4$}
\psfrag{Nr=4AAABBB, Ne=8AAAA}[l][l][0.26]{$N_\rr=4, N_\ee=8$}
%\psfrag{Nt=256AAAB}[l][l][0.2]{$M=256$}

%y-axis

\psfrag{20}[c][b][0.25]{$20$}
\psfrag{80}[c][b][0.25]{$80$}
\psfrag{40}[c][b][0.25]{$40$}
\psfrag{100}[c][b][0.25]{$100$}
\psfrag{60}[c][b][0.25]{$60$}
\psfrag{120}[c][b][0.25]{$120$}

%x-axis
\psfrag{0}[r][c][0.25]{$0$}
\psfrag{10}[r][c][0.25]{$10$}
\psfrag{15}[r][c][0.25]{$15$}
\psfrag{5}[r][c][0.25]{$5$}
\psfrag{21}[r][c][0.25]{$20$}
}}
\vspace*{-7mm}\caption{The ergodic secrecy rate as a function of $L$ for $M=128$, $\log \rho_\mm=0$ dB and $\log \rho_\ee=-10$ dB. The solid lines and the circles which follows each other are the approximation in \eqref{eq:asy} and numerical simulations, respectively.\vspace*{-3mm}}
\label{fig:3}
\end{figure}

\begin{figure}[t]
\hspace*{-0.75cm}  
%%\centering
\resizebox{1.1\linewidth}{!}{
\pstool[width=.35\linewidth]{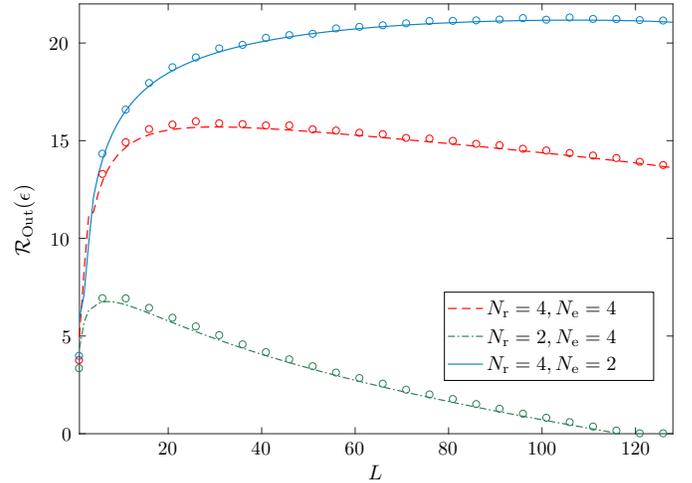}{
\psfrag{Outage Probability}[c][t][0.27]{$\mar_\out(\epsilon)$}
\psfrag{Selected Antenna}[c][c][0.27]{$L$}
\psfrag{Nr=2AAABBB, Ne=4AAAAA}[l][l][0.26]{$N_\rr=2, N_\ee=4$}
\psfrag{Nr=4AAABBB, Ne=4AAAAA}[l][l][0.26]{$N_\rr=4, N_\ee=4$}
\psfrag{Nr=4AAABBB, Ne=2AAAAA}[l][l][0.26]{$N_\rr=4, N_\ee=2$}
%\psfrag{Nt=256AAAB}[l][l][0.2]{$M=256$}

\psfrag{20}[c][b][0.25]{$20$}
\psfrag{80}[c][b][0.25]{$80$}
\psfrag{40}[c][b][0.25]{$40$}
\psfrag{100}[c][b][0.25]{$100$}
\psfrag{60}[c][b][0.25]{$60$}
\psfrag{120}[c][b][0.25]{$120$}

%x-axis
\psfrag{0}[r][c][0.25]{$0$}
\psfrag{10}[r][c][0.25]{$10$}
\psfrag{15}[r][c][0.25]{$15$}
\psfrag{5}[r][c][0.25]{$5$}
\psfrag{21}[r][c][0.25]{$20$}
}}
\vspace*{-7mm}\caption{The $\epsilon$-outage secrecy rate for $\epsilon=0.01$ versus $L$ for $M=128$, $\log \rho_\mm=0$ dB and $\log \rho_\ee=-10$ dB. The solid lines and the circles following each other closely are the approximation and numerical simulations, respectively.\vspace*{-3.2mm}}
\label{fig:out-L}
%\vspace{-4.0mm}
\end{figure}

In this section, we investigate the impacts of \ac{tas}~on~the~secrecy performance in both Scenarios A and B. Let us start with Scenario~A. As it was discussed, the secrecy performance in this case is characterized by the ergodic secrecy rate whose large-system approximation is given in Section~\ref{sec:active}. Considering the ergodic secrecy rate $\mar_\rmE (\mas)$ as a function of $L$, one observes that for different choices of $\rho_\ee$, $\rho_{\rm m}$, $N_{\rm r}$ and $N_\ee$, the ergodic secrecy rate may strictly increase with $L$ within the interval $\set{1, \ldots, M}$ or have a maxima at some integer $L^\star < M$. This observation suggests that for the considered setting the secrecy performance can be enhanced in some cases via \ac{tas}. Fig.~\ref{fig:3} illustrates this point. In this figure, the ergodic secrecy rate is plotted as a function of $L$, for several realizations of the setting with $M=128$ considering both the large-system approximation and numerical simulations. The \ac{snr}s at the legitimate receiver and eavesdropper are set to $\log \rho_\mm=0$ dB and $\log \rho_\ee=-10$ dB, respectively. As the figure shows, the ergodic secrecy rate in some curves meets its maximum at some values of $L$ which is significantly smaller than $M$. This observation depicts that \ac{tas} in these scenarios, not only benefits in terms of \ac{rf}-cost and complexity, but also enhances the secrecy performance of the system. The intuition behind this behavior comes from the fact that the growth in the number of selected antennas improves the quality of the both channels. For some cases, including those shown in Fig.~\ref{fig:3}, the improvement from the eavesdropper's point of view dominates the overall growth in the secrecy rate, if a certain number of active antennas is surpassed. This means that by setting $L$ to be more than this given number, the quality improvement at the eavesdropping terminal starts to exceed the enhancement at the legitimate receiver. Considering Scenario~B, similar behavior can be observed in terms of $\epsilon$-outage secrecy rate $\mar_\out(\epsilon)$. In Fig.~\ref{fig:out-L} the $\epsilon$-outage secrecy rate for $\epsilon=0.01$ has been plotted in terms of $L$ for several examples considering $M=128$, $\log \rho_\mm=0$ dB and $\log \rho_\ee=-10$ dB. %Similar to Scenario~A, the secrecy performance in this case can be enhanced via \ac{tas}.

\subsection{Characterization of Secrecy Enhancement}
Based on the latter observations, one may intuitively state that \ac{tas} plays a constructive role on the secrecy performance when the eavesdropping terminal starts to experience prevailing improvements in its channel quality by growth in $L$ at some $L<M$. The characterization of the settings in which this behavior is observed is however not trivial,~since~the~performance metrics in general depend on several~parameters.~In the sequel, we invoke our large-system results to characterize these settings. %set of system parameters in which the secrecy performance of the system is enhanced via \ac{tas}. compared to the case of full complexity, i.e., $L=M$. On the other hand, for those cases in which the legitimate receiver observes better conditions for all choices of $L\in{\left\lbrace  1, \ldots, M\right\rbrace }$, the \ac{tas} protocol $\mas$ does not improve the performance compared to full \ac{tas} and the optimal choice in this sense is $L=M$.
For this aim, we first define the~``prevalence~set'' for the legitimate receiver and the eavesdropper.

\begin{definition}[Prevalence Set]
\label{def:strong}
%Consider the setup illustrated in Section~\ref{sec:sys}. Let the performance of the system for $L$ active transmit antennas, selected by the protocol $\mas$, be characterized by the metric 
%%with $N_\rr$ receive antennas and \ac{snr} $\rho_\mm$ 
%than an eavesdropper with $N_\ee$ antennas and \ac{snr} $\rho_\ee$ in the asymptotic regime %than the eavesdropper,
%with $N_\ee$ antennas and \ac{snr} $\rho_\ee$ is said to be relatively stronger than a legitimate receiver with $N_\rr$ antennas and \ac{snr} $\rho_\mm$ in the asymptotic regime 
Let $\mam(L)$ denote the secrecy performance metric for $L$ active transmit antennas. The legitimate receiver is said to be relatively prevailing, if $\mam(L)$ is a monotonically increasing function of $L$. Moreover, the set of all tuples $(\rho_\mm, \rho_\ee, N_\rr, N_\ee)$ for which the legitimate receiver is relatively prevailing is referred to as the prevalence set for the legitimate receiver represented by $\setS_\rec$. Similarly, the eavesdropper is said to be relatively prevailing if
\begin{align}
\min\set{ \argmax_{L\in\set{1,\ldots,M}} \mam(L)} < M.
\end{align}
The prevalence set for the eavesdropper $\setS_\eve$ is then defined as the set of all tuples $(\rho_\mm, \rho_\ee, N_\rr, N_\ee)$ for which the eavesdrop-per is relatively prevailing.%in the asymptotic regime is relatively stronger than the~legitimate~receiver.
\end{definition}

Definition~\ref{def:strong} partitions the realizations of the setting into two sets. In the former set, represented by $\setS_\rec$, the growth in the number of active antennas improves the communication quality over the main channel always more than over the eavesdropper channel. The latter set, denoted by $\setS_\eve$, moreover, encloses the settings in which the improvement at the eavesdropper channel starts to prevail when $L$ exceeds at some $L<M$. Consequently, the secrecy performance in this case is enhanced by employing the protocol~$\mas$. %Considering the discussions in Sections~\ref{sec:active} and \ref{sec:passive}, the prevalence set for the eavesdropper and the legitimate receiver can be determined by investigating the asymptotic terms of the ergodic and the $\epsilon$-outage secrecy rate, respectively. Therefore, in the sense of secrecy performance, the optimal choice for $L$ in this case is $L^\star=M$, and employing the \ac{tas} protocol $\mas$ only benefits in terms of \ac{rf}-cost and complexity.
%In the sequel, we derive some sufficient conditions for prevalence in both Scenarios~A and B. %For cases in which the eavesdropper is relatively prevailing, we further employ the large-system approximations to analytically derive the optimal number of active transmit~antennas. 

\subsection{Sufficient Conditions for Prevalence}
%In Scenario~A, the ergodic secrecy rate is considered to be the secrecy performance metric, i.e., $\mam(L)=\mar_\rmE(\mas)$. The secrecy performance in Scenario B is quantified with $\epsilon$-outage secrecy rate which means that $\mam(L)=\mar_\out(\epsilon)$. 
Using the large-system approximation, one can determine $\setS_{\rm R}$ and $\setS_{\rm E}$ for large $M$ analytically. The result is however of a complicated form in general. Alternatively, one may derive a set of sufficient conditions for which the prevalence of the legitimate or eavesdropping terminal is guaranteed. Theorem~\ref{thm:2} gives a set of sufficient conditions for the legitimate receiver to be relatively prevailing.
%Consequently, the relative strength sets of the eavesdropper and legitimate receiver are derived as in the following.

\begin{theorem}
\label{thm:2}
Let the transmitter be equipped with $M$ transmit antennas and assume the asymmetrically asymptotic~regime~of eavesdropping. For a given tuple $T=(\rho_\mm, \rho_\ee, N_\rr, N_\ee)$, define the fixed-point function $f(\cdot|T)$~to~be
\begin{align}
f(\ell | T)\coloneqq F(\ell) + f_{\rm R}(\ell| \rho_\mm, N_\rr) - f_{\rm E}(\ell| \rho_\ee, N_\ee) \label{eq:f_end}
\end{align}
where the function $F(\ell)$ reads
\begin{align}
F(\ell) = \frac{\rho_\mm u \hspace*{.7mm} U_\mm }{U_\mm+\rho_\mm \eta_t} \prnt{ 1+ 2 \Lambda(\ell) \frac{U_\mm \prnt{U_\mm -1}}{\rho_\mm \eta_t V_\mm} }
\end{align}
with $u$ and $\eta_t$ being defined in Theorem~\ref{thm:1}, $U_\mm=\min\set{\ell , N_\rr}$, $V_\mm=\max\set{\ell , N_\rr}$ and
\begin{align}
\Lambda(\ell) = \frac{\psi}{2} \prnt{\frac{\rho_\mm \eta_t }{U_\mm+\rho_\mm \eta_t}}^2.
\end{align}
Moreover, $f_{\rm R}(\ell| \rho_\mm, N_\rr)$ and $f_{\rm E}(\ell| \rho_\ee, N_\ee)$ are given by \eqref{eq:f_r} and \eqref{eq:f_e} on the top of the~next~page with $E(\ell)$ reading
\begin{align}
\hspace*{-1mm}E(\ell) \hspace*{-.7mm}=\hspace*{-.7mm} 1 \hspace*{-.7mm}-\hspace*{-.7mm} \frac{U_\mm }{U_\mm \hspace*{-.7mm} + \hspace*{-.7mm} \rho_\mm \eta_t} \left[ 1\hspace*{-.7mm}-\hspace*{-.7mm} \Lambda(\ell) \frac{U_\mm \hspace*{-.7mm}+ \hspace*{-.7mm}\prnt{2 U_\mm \hspace*{-.7mm} - \hspace*{-.7mm} 1} \rho_\mm \eta_t}{U_\mm V_\mm} \right].
\end{align}
Then, the legitimate receiver is relatively prevailing in both Scenarios~A and B if $f(\ell| T) > 0$ for all real  $\ell\in [1,M] $. 
\begin{figure*}[!t]
\begin{subequations}
\begin{align}
f_{\rm R}(\ell| \rho_\mm, N_\rr) &\coloneqq \mone\set{\ell<N_\rr} \left[ \log\prnt{1+\frac{\rho_\mm \eta_t}{\ell}}-E(\ell) \right] + \mone\set{\ell>N_\rr} \frac{N_\rr(N_\rr-1) \Lambda(\ell)}{ \ell^2}, \label{eq:f_r} \\
f_{\rm E}(\ell| \rho_\ee, N_\ee)&\coloneqq \mone\set{\ell<N_\ee} \log\prnt{1+ \rho_\ee N_\ee} + \mone\set{\ell>N_\ee} \frac{\rho_\ee N_\ee }{1+\rho_\ee \ell} \label{eq:f_e}
\end{align}
\end{subequations}
\centering\rule{17cm}{0.1pt}
%\hrulefill
%
\vspace*{2pt}
\end{figure*}
\end{theorem}

\begin{prf}
The proof follows bounding the first~derivatives~of~the large-system approximations for the ergodic and~$\epsilon$-outage~secrecy rate by a similar term, and is given in~Appendix~\ref{app:2}.
\end{prf}
%One may note that 

%\subsubsection*{Passive Eavesdropping}
% For this case, one can determine the relative strength sets $\setS_\rec $ and $\setS_\eve $ as given in Theorem~\ref{thm:3}.
%
%\begin{theorem}
%\label{thm:3}
%Consider the \ac{mimome} channel in Section~\ref{sec:sys} with $M$ transmit antennas, and suppose that the $\epsilon$-outage secrecy rate is the performance metric. Let the main channel be overheard the channel in a relatively asymptotic regime of eavesdropping as defined in Definition~\ref{def:rel}. Define the fixed-point function $f_{\rm P}(\cdot|T)$ to be
%\begin{align}
%\textcolor{red}{
%f_{\rm P}(L| T)
%}
%\end{align}
%for a given tuple $T=(\rho_\mm, \rho_\ee, N_\rr, N_\ee)$. Then, $ T \in \setS_\rec $ if  $f_{\rm P}(L| T)>0$ for all $ L \in \set{1,\ldots,M} $ and $ T \in \setS_\eve $ if there exist at least a real scalar $L_0\in [0,M] $ for which $f_{\rm P}(L_0| T)=0$.
%\end{theorem}
%
%\begin{prf}
%proof
%\end{prf}

Theorem~\ref{thm:2} intuitively indicates that the legitimate receiver is prevailing when the growth in the achievable rate over the main channel by increasing the number of active antennas always dominates the growth over the eavesdropper channel. In fact, the first two terms in the right hand side of \eqref{eq:f_end} bound the rate growth over the main channel while $f_{\rm E}(\ell| \rho_\ee, N_\ee)$ describes the improvement in the quality of the eavesdropper channel in the large-system limit. Using Theorem~\ref{thm:2}, one can discuss whether secrecy enhancement is achievable in the setting via \ac{tas} or not. One should note that this theorem states only a sufficient condition. This means that there exist tuples which do not fulfill the conditions given in Theorem~\ref{thm:2} and still are optimal under full complexity in the sense of secrecy performance. For these cases, one may further study necessary conditions. In the following, we study~some~examples.
%one may conclude whether employing a subset of transmit antennas can additionally improve secrecy performance or only reduce the \ac{rf}-costs and complexity of the system. We further investigate the applications of these result in the following example.
\begin{example}
\label{ex:0}
Consider the following two scenarios:
\begin{itemize}
\item[(a)] The legitimate and eavesdropping terminals are equipped with $N_\rr=8$ and $N_\ee=2$ antennas, respectively and we have $\log \rho_\mm=0$ dB, $\log \rho_\ee=-10$ dB and $M=128$.
\item[(b)] The eavesdropper is equipped with a single antenna while $N_\rr>1$. The number of transmit antennas unboundedly grows large, i.e., $M \uparrow \infty$.
\end{itemize}
%Considering Example , it is straightforward to show that the legitimate receiver is relatively prevailing. In fact u
From Theorem~\ref{thm:2}, one can show that for the setting in (a) the sufficient conditions are satisfied, and thus, the legitimate receiver is relatively prevailing in both Scenarios~A and B. This result agrees with this intuition that the legitimate receiver is prevailing, since both the number of receive antennas and the \ac{snr} are relatively better at the this terminal. %As Fig.~\ref{fig:Ex-1} shows for this case the secrecy performance is maximized when $L = M$.

For (b), we invoke Theorem~\ref{thm:2} and derive a set of conditions under which the legitimate receiver becomes relatively prevailing. Since $M \uparrow \infty$, one can show that for this case $\Lambda(\ell)\approx \psi /2$ and $E(\ell)\approx 1$. Moreover, the function $F(\ell)$ in the large-system limit can be approximated as
\begin{align}
F(\ell) \approx \frac{u \hspace*{.7mm} U_\mm }{N_\rr \ell + N_\rr M f_{N_\rr +1}(u)}
\end{align}
with $u$ and $f_{N_\rr+1}(u)$ given in Theorem~\ref{thm:1}. Substituting in~\eqref{eq:f_end}, the constraints in \eqref{eq:LeqN} at the top of the next page is derived. When $M \uparrow \infty$, one concludes that $N_\rr \geq 1+\sqrt{2M}$~is~sufficient for the prevalence of the legitimate receiver. Note that~this~co-nstraint does not depend on the \ac{snr}s. For instance, considering $M=128$, a legitimate terminal with $N_\rr=17$ antennas is relatively prevailing for any choice of $\rho_\mm$ and $\rho_\ee$. Fig. \ref{fig:ex-22} shows the achievable ergodic rate for $M=128$, $N_\ee=1$ and $N_\rr=17$ considering several choices of $\rho_\ee$ and $\rho_\mm$. As the figure depicts, the optimal choice for $L$ in all the cases is $L^\star=M$ which agrees with the analytic result.
% for $\ell<N_\rr$ we have $f_{\ell}(\cdot|T) > 0$. As $\ell$ takes values greater than $N_\rr$, the constraint in \eqref{eq:f_end} reduces to
%\begin{align}
%N_\rr(N_\rr-1) > \frac{2 \rho_\ee \ell^2 }{\prnt{1+\rho_\ee \ell}\psi}
%\end{align}
%for $\ell\in[N_\rr,M]$. As the right hand side of the inequality is an increasing function of $\ell$, it is sufficient that we have
%\begin{align}
%N_\rr(N_\rr-1) > \frac{2 \rho_\ee M^2 }{\prnt{1+\rho_\ee M}\psi}
%\end{align}
%Considering the case in which $\rho_\ee = -10$ dB and $M=128$, one can guarantee that the legitimate terminal is relatively stronger than the eavesdropper when $N_\rr\geq 13$. \textcolor{red}{continue}
\end{example}

\begin{figure*}[!t]
\begin{align}
\begin{cases}
 \dfrac{u \ell }{N_\rr \ell + N_\rr M f_{N_\rr +1}(u)} + \log \prnt{1+\dfrac{\rho_\mm \eta_t}{\ell}}  \geq 1+ \dfrac{\rho_\ee }{ 1+\rho_\ee \ell} \qquad &\ell < N_\rr \vspace*{3mm} \\
\dfrac{u }{\ell + M f_{N_\rr +1}(u)} + \dfrac{N_\rr(N_\rr-1)}{2 \ell^2} \psi   \geq  \dfrac{\rho_\ee }{ 1+\rho_\ee \ell} &\ell > N_\rr. \vspace*{2mm}
\end{cases} \label{eq:LeqN}
\end{align}
\centering\rule{17cm}{0.1pt}
%\hrulefill
%
\vspace*{1pt}
\end{figure*}
\section{Optimal Number of Active Antennas}
\label{sec:optimal}
When the eavesdropper is relatively prevailing, the secrecy performance metric is maximized by choosing the number of active antennas optimally. We investigate this problem through some examples considering both Scenarios~A and B.
%In this case, one can utilize the large-system approximation of the performance metrics to analytically derive the optimal number of active transmit antennas.
\subsection{Scenario A}
The large-system approximation of $\mar_\rmE \prnt{\mas}$ in \eqref{eq:asy}~is~a~fu-nction of $L$ whose maxima occurs at some~$L^\star \in [1:M]$~when the eavesdropper is relatively prevailing. We derive this maxima analytically for some examples in the sequel.
%In this case, the results given in Section \ref{sec:result} can be employed, in order to find the optimal number of selected transmit antennas. More precisely, using \eqref{eq:eta_final} and \eqref{eq:sigma_final}, one can determine the ergodic secrecy rate and secrecy outage probability in \eqref{eq:asy} and \eqref{eq:out} as functions of $L$ whose maximizers are found either analytic or via a linear search. We investigate this problem further through the following examples by considering the ergodic secrecy rate as the measure. Same discussions can also be considered for the secrecy outage probability
\begin{example}[Single-antenna receivers]
\label{ex:1}
Consider the scenario in which the receiving terminals are equipped with a single antenna, i.e., $N_\rr=N_\ee=1$. Assume that the eavesdropper's \ac{csi} is available at the transmitter.

We intend to derive the optimal number of active antennas $L^\star$ which maximizes $\mar_\rmE\prnt{\mas}$. To do so, we initially assume that with $L^\star$ active transmit antennas the setting performs in the asymmetrically asymptotic regime of eavesdropping, i.e., $L^\star \gg 1$. We later show that this prior assumption is true. By substituting into \eqref{eq:eta_final} and \eqref{eq:sigma_final}, $\eta$ and $\sigma^2$ are determined as
\begin{figure}[t]
\hspace*{-0.75cm}  
%%\centering
\resizebox{1.05\linewidth}{!}{
\pstool[width=.35\linewidth]{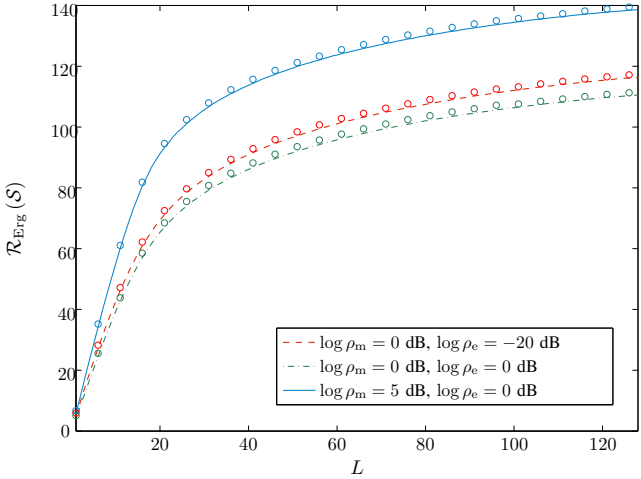}{
\psfrag{RE}[c][t][0.27]{$\mar_\rmE\prnt{\mas}$}
\psfrag{Lt}[c][c][0.27]{$L$}
\psfrag{rhom=0AAA-SABA,rhoe=0AAAA-SABAQ}[l][l][0.24]{$\log \rho_\mm=0$ dB, $\log \rho_\ee=0$ dB}
\psfrag{rhom=0AAA-SABA,rhoe=-20AAA-SABAQ}[l][l][0.24]{$\log \rho_\mm=0$ dB, $\log \rho_\ee=-20$ dB}
\psfrag{rhom=5AAA-SABA,rhoe=0AAAA-SABAQ}[l][l][0.24]{$\log \rho_\mm=5$ dB, $\log \rho_\ee=0$ dB}
%\psfrag{Nt=256AAAB}[l][l][0.2]{$M=256$}

%y-axis
\psfrag{0}[r][c][0.25]{$0$}
\psfrag{20x}[r][c][0.25]{$20$}
\psfrag{80x}[r][c][0.25]{$80$}
\psfrag{40x}[r][c][0.25]{$40$}
\psfrag{100x}[r][c][0.25]{$100$}
\psfrag{60x}[r][c][0.25]{$60$}
\psfrag{120x}[r][c][0.25]{$120$}
\psfrag{140x}[r][c][0.25]{$140$}

%x-axis
\psfrag{20}[c][b][0.25]{$20$}
\psfrag{40}[c][b][0.25]{$40$}
\psfrag{60}[c][b][0.25]{$60$}
\psfrag{80}[c][b][0.25]{$80$}
\psfrag{120}[c][b][0.25]{$120$}
\psfrag{140}[c][b][0.25]{$140$}
\psfrag{100}[c][b][0.25]{$100$}
}}
\caption{
%The ergodic secrecy rate versus $L$ for different setups.  The curves have been plotted for $M=128$, $N_\ee=1$ and $N_\rr=17$. All the curves suggest that the ergodic secrecy rate increases monotonically as the number of used antennas increases. The ergodic secrecy rate is maximized when the number of selected antennas equals to the number of total available transmit antennas. The solid lines indicate the approximated ergodic secrecy rate and the numerical simulations are plotted via the circles.
The ergodic secrecy rate versus $L$ for $M=128$, $N_\ee=1$ and $N_\rr=17$. 
%All the curves suggest that the ergodic secrecy rate increases monotonically as the number of used antennas increases. The ergodic secrecy rate is maximized when the number of selected antennas equals to the number of total available transmit antennas. 
The solid lines indicate the approximated ergodic secrecy rate and the numerical simulations are plotted via the circles.\vspace*{-4mm}
}
\label{fig:ex-22}
%\vspace{-4.0mm}
\end{figure}
\begin{subequations}
\begin{align}
\eta&=\log \prnt{\frac{1+\rho_\mm L\prnt{1+ \loge{M L^{-1}}}}{1+\rho_\ee L}}
\end{align}
\begin{align}
\sigma^2&= \left[ \frac{\rho_\mm^2 \hspace*{1mm} L \left( 2- L M^{-1} \right)}{\prnt{1+\rho_\ee L}^2} + \frac{L \rho_\ee^2}{(1+\rho_\ee L)^2} \right] \psi^2.
\end{align}
\end{subequations}
%in which the ergodic secrecy rate meets its maximum with $L=18$ active transmit antennas. 
\end{example}
Under the assumption $L^\star\gg 1$, the achievable ergodic secrecy rate for $M\uparrow \infty$ is further approximated as $\mar_\rmE\prnt{\mas}\approx \eta$. To find $L^\star$, we define the function
\begin{align}
\mar(\ell) \coloneqq \log \prnt{\frac{1+\rho_\mm \ell+\rho_\mm \ell \hspace*{.5mm} \loge{M \ell^{-1}}}{1+\rho_\ee \ell}}
\end{align}
for real $\ell$. $\mar(\cdot)$ is the real envelope of the ergodic secrecy~rate whose values at integer points give the ergodic secrecy rate for the given number of active antennas. In this case, one can straightforwardly show that for any choice of $\rho_\mm$ and $\rho_\ee\neq 0$, there exists some real $\ell\in[1,M]$ for which $\mar'(\ell)<0$. This fact indicates that with $N_\rr=N_\ee=1$, the eavesdropper is relatively prevailing\footnote{Note that $\mar'(\ell)<0$ holds for large $\ell$ which agrees with the initial assumption of being in the asymmetrically asymptotic regime of eavesdropping.} as long as $\rho_\ee \neq 0$, and thus, $L^\star <M$. To find $L^\star$, one notes that $\mar''(\ell) \leq 0$ for $ \ell \in[0,M]$, and thus, $L^\star$ is the closest integer to the maxima of $\mar(\cdot)$. Consequently, the optimal number of active transmit antennas is approximated as $L^\star \approx \lfloor \ell^\star\rceil$ where $\ell^\star$ satisfies
\begin{align}
\rho_\ee \ell^\star + \loge {\ell^\star} + \frac{\rho_\ee}{\rho_\mm}= \loge{M}. \label{eq:L_opt}
\end{align}
\begin{figure}[t]
\hspace*{-0.75cm}  
%%\centering
\resizebox{1.15\linewidth}{!}{
\pstool[width=.35\linewidth]{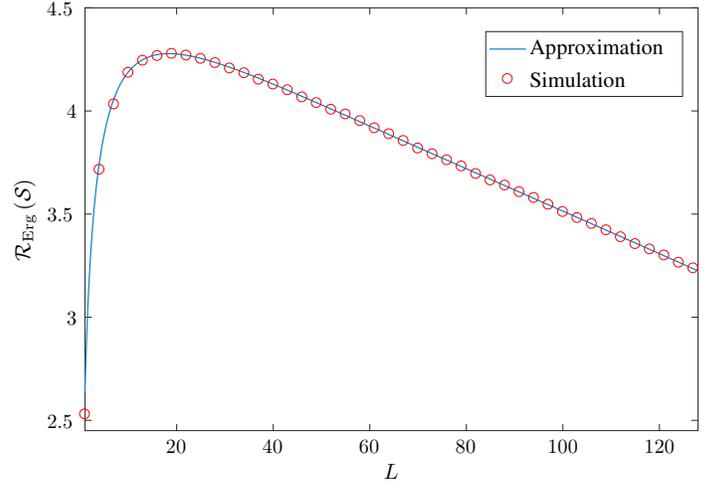}{
\psfrag{Rerg}[c][t][0.27]{$\mar_\rmE\prnt{\mas}$}
\psfrag{Lt}[c][c][0.27]{$L$}
\psfrag{Nr=1-Ne=1-S=128AAAB}[l][l][0.27]{Approximation}
\psfrag{Nr=1-Ne=1-R=128AAAB}[l][l][0.27]{Simulation}

%y-axis
\psfrag{20}[c][b][0.25]{$20$}
\psfrag{40}[c][b][0.25]{$40$}
\psfrag{60}[c][b][0.25]{$60$}
\psfrag{80}[c][b][0.25]{$80$}
\psfrag{100}[c][b][0.25]{$100$}
\psfrag{120}[c][b][0.25]{$120$}

%x-axis
\psfrag{2.5}[r][c][0.25]{$2.5$}
\psfrag{3}[r][c][0.25]{$3$}
\psfrag{3.5}[r][c][0.25]{$3.5$}
\psfrag{4}[r][c][0.25]{$4$}
\psfrag{4.5}[r][c][0.25]{$4.5$}
}}
\vspace*{-8mm}\caption{
%$\mar_\rmE(\mas)$ in Example~\ref{ex:1} in terms of the number of selected antennas. The solid line and the circles show the approximation given by \eqref{eq:asy} and numerical simulations respectively considering $M=128$, $\rho_\mm=0$ dB and $\rho_\ee=-10$ dB. As it is observed, $L^\star=18$ is suggested by both the approximation and simulation results.
$\mar_\rmE(\mas)$ in Example~\ref{ex:1} in terms of $L$. The solid line and the circles show the approximation given by \eqref{eq:asy} and numerical simulations respectively for $M=128$, $\log \rho_\mm=0$ dB and $\log \rho_\ee=-10$ dB. As it is observed, $L^\star=18$ is suggested by both the approximation and simulation results.
}
\label{fig:4}
%\vspace{-4.0mm}
\end{figure}
From \eqref{eq:L_opt}, one observes that $L^\star$ grows with $M$, and therefore, the eavesdropping regime is asymmetrically asymptotic, i.e., the initial assumption $L^\star \gg 1$ holds. Moreover, by reducing $\rho_\ee\downarrow 0$ in \eqref{eq:L_opt}, $L^\star=M$ which agrees with the fact that in the absence of eavesdroppers, the achievable ergodic secrecy rate is a monotonically increasing function of $L$. Fig.~\ref{fig:4} shows the ergodic secrecy rate as a function of $L$ for $\log \rho_\ee=-10$ dB and $\log \rho_\mm=0$ dB assuming that $M=128$ antennas are available at the transmit side. By solving the fixed-point equation in \eqref{eq:L_opt}, $\ell^\star=18.4$ is obtained which results in $L^\star=18$. This result is confirmed by numerical simulations in Fig.~\ref{fig:4}.

\begin{example}[Multi-antenna eavesdropper]
\label{ex:2}
Consider a scenario with a single antenna legitimate receiver whose channel is being overheard by a sophisticated multi-antenna terminal, i.e., $N_\rr=1$ and $N_\ee$ growing large. Assume that the \ac{bs} knows the \ac{csi} of the eavesdropper.

%As $N_\ee$ is large, the eavesdropper overhears the channels in an asymmetrically asymptotic regime of eavesdropping as denoted in Definition~\ref{def:rel}. Using 
From Theorem~\ref{thm:1}, $\eta$ and $\sigma^2$ are given by% in \eqref{eq:eta_final} and \eqref{eq:sigma_final} are derived as
\begin{subequations}
\begin{align}
\eta&=\log \prnt{\frac{1+\rho_\mm L\prnt{1+ \loge{M L^{-1}}}}{(1+\rho_\ee N_\ee)^{L}}} \label{eq:ex-21} \\
\sigma^2&= \frac{\rho_\mm^2 \psi^2 \hspace*{1mm} L \left( 2- L M^{-1} \right)}{\prnt{1+\rho_\ee L}^2} + \frac{\psi^2}{L}. \label{eq:ex-22}
\end{align}
\end{subequations}
%Using Theorem~\ref{thm:2} one can show that there exist a set of pairs $(\rho_\mm , \rho_\ee)$ over which the eavesdropper becomes relatively stronger than the legitimate terminal. To find the optimal number of active transmit antennas $L^\star$ for this set,
In contrast to Example \ref{ex:1}, $\mar_\rmE\prnt{\mas}$ in this example can not be approximated by $\eta$, since  $\xi=\eta/\sigma$ is not necessarily large. Consequently, we employ \eqref{eq:asy} to accurately approximate the achievable ergodic rate. Define $\mar(\cdot)$ over the real axis as
\begin{align}
\mar(\ell) \coloneqq s(\ell)\phi\prnt{\frac{f(\ell)}{s(\ell)}} + f(x) \rmQ\prnt{-\frac{f(\ell)}{s(\ell)}}
\end{align}
%\begin{align}
%f(x) \coloneqq \log \prnt{\frac{1+\rho_\mm x+\rho_\mm x \hspace*{.5mm} \loge{M x^{-1}}}{1+\rho_\ee x}}
%\end{align}
where $f(\ell)$ and $s(\ell)$ are given by
\begin{subequations}
\begin{align}
f(\ell) &= \log \prnt{\frac{1+\rho_\mm \ell \prnt{1+ \loge{\frac{M}{\ell}}}}{(1+\rho_\ee N_\ee)^{\ell}}} \\
s(\ell) &= \sqrt{\frac{\rho_\mm^2 \psi^2 \hspace*{1mm} \ell \left( 2- \frac{\ell}{M} \right)}{\prnt{1+\rho_\ee \ell}^2} + \frac{\psi^2}{\ell} }.
\end{align}
\end{subequations}
With similar lines of inference as in Example~\ref{ex:1}, one concludes that for any non-zero choices of $\rho_\ee$ and $\rho_\mm$ the eavesdropper is relatively prevailing. This result is intuitive, since the eavesdropper is more sophisticated compared to the one considered in Example~\ref{ex:1}. Consequently, $L^\star \approx \lfloor \ell^\star\rceil$ where $\ell^\star$ satisfies\footnote{One may show that the fixed-point equation in \eqref{eq:L2} has always a solution within the interval of $[1,M]$ which implies the fact that the eavesdropper is relatively prevailing for any non-zero choices of $\rho_\ee$ and $\rho_\mm$. }
\begin{figure}[t]
\hspace*{-0.75cm}  
%%\centering
\resizebox{1.1\linewidth}{!}{
\pstool[width=.35\linewidth]{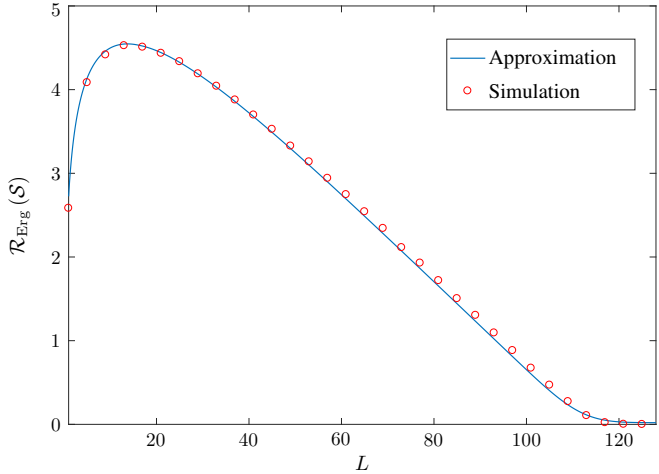}{
\psfrag{Rerg}[c][t][0.27]{$\mar_\rmE\prnt{\mas}$}
\psfrag{Lt}[c][c][0.27]{$L$}
\psfrag{Nr=Ne=16Nt=SAAAABB}[l][l][0.27]{Approximation}
\psfrag{Nr=Ne=16Nt=RAAAABB}[l][l][0.27]{Simulation}

%y-axis
\psfrag{20}[c][b][0.25]{$20$}
\psfrag{40}[c][b][0.25]{$40$}
\psfrag{60}[c][b][0.25]{$60$}
\psfrag{80}[c][b][0.25]{$80$}
\psfrag{100}[c][b][0.25]{$100$}
\psfrag{120}[c][b][0.25]{$120$}

%x-axis
\psfrag{0}[r][c][0.25]{$0$}
\psfrag{1}[r][c][0.25]{$1$}
\psfrag{2}[r][c][0.25]{$2$}
\psfrag{3}[r][c][0.25]{$3$}
\psfrag{4}[r][c][0.25]{$4$}
\psfrag{5}[r][b][0.25]{$5$}
}}
\vspace{-8mm}\caption{
%The ergodic secrecy rate in Example~\ref{ex:2} vs. $L$ for the case with $N_\ee=16$, $M=128$, $\log \rho_\mm=0$ dB and $\log \rho_\ee=-25$ dB. The approximation and numerical simulations are denoted by the solid line and circles respectively, and both suggest $L^\star=14$.
The ergodic secrecy rate in Example~\ref{ex:2} versus $L$ for $M=128$, $N_\ee=16$, $\log \rho_\mm=0$ dB and $\log \rho_\ee=-25$ dB. Both the approximation and simulations, denoted respectively by the solid line and circles, suggest $L^\star=14$.
}
\label{fig:5}
\end{figure}
\begin{align}
h(\ell^\star) &\phi\prnt{h(\ell^\star)} \frac{f'(\ell^\star)s(\ell^\star)-f(\ell^\star)s'(\ell^\star)}{s(\ell^\star)} 
%- \frac{1}{2} s'(\ell^\star)} 
\nonumber \\
&= \frac{1}{2} f'(\ell^\star) \rmQ\prnt{-h(\ell^\star)} +\frac{1}{2} s'(\ell^\star) \phi\prnt{h(\ell^\star)} . \label{eq:L2}
\end{align}
%For the choices of $(\rho_\mm, \rho_\ee)$ in which the fixed point equation in \eqref{eq:L2} has no solution in $[0,M]$, one can see that $f_{\rmA}(\ell|T)$ for the given tuple $T$ is positive within $[0,M]$, and thus, due to Theorem~\ref{thm:2}, the legitimate terminal is relatively stronger than the eavesdropper meaning that $L^\star=M$.
with $h(\ell)=f(\ell)/s(\ell)$. In Fig.~\ref{fig:5}, $\mar_\rmE(\mas)$ is sketched versus $L$ for $N_\ee=16$ assuming $\log \rho_\ee=-25$ dB, $\log \rho_\mm=0$ dB and $M=128$. From \eqref{eq:L2}, the maxima of the function $\mar(\ell)$ is derived as $\ell^\star=13.7$ which recovers $L^\star=14$ given by simulations.
\end{example}

\subsection{Scenario B}
Considering Scenario B, a similar approach can be taken to derive the optimal number of active antennas. We investigate this case through the following example.
% using the large-system approximation given in \eqref{eq:R_out-Final}.
\begin{example}[Passive eavesdropping]
\label{ex:3}
Similar to Example~\ref{ex:1}, consider a case with $N_\rr=N_\ee=1$. Let $\rho_\ee=\rho_\mm=\rho$, and assume that the eavesdropper's \ac{csi} is not available at the \ac{bs}.

The performance metric is the $\epsilon$-outage secrecy rate which in the large-system is approximated by \eqref{eq:R_out-Final} with $\eta$ and $\sigma$ %given by
\begin{subequations}
\begin{align}
\eta&=\log \prnt{\frac{1+\rho L\prnt{1+ \loge{M L^{-1}}}}{1+\rho L}}\\
\sigma^2&=  \left[ \frac{\rho \psi }{1+\rho L}\right]^2 \prnt{3-\frac{L}{M}}.
\end{align}
\end{subequations}
In order to investigate the prevalence, we define %the function
\begin{align}
\mar(\ell) \hspace*{-.7mm}\coloneqq \hspace*{-.7mm}\log \prnt{\frac{1\hspace*{-.7mm}+\hspace*{-.7mm}\rho \ell\hspace*{-.7mm}+\hspace*{-.7mm}\rho \ell  \loge{M \ell^{-1}}}{1+\rho \ell}} \hspace*{-.7mm}+\hspace*{-.7mm}   \frac{\rho \psi q_0 }{1+\rho \ell}  \sqrt{3\hspace*{-.7mm}-\hspace*{-.7mm}\frac{\ell}{M}}
\end{align}
where $q_0=\rmQ^{-1}\prnt{1-\epsilon}$. It is then trivial to show that 
\begin{align}
\mar'(\ell) \coloneqq &\frac{ \rho \loge\prnt{M \ell^{-1}} - \rho (1+\rho \ell) }{\prnt{1+\rho \ell} \prnt{1+\rho \ell+\rho \ell \loge\prnt{M \ell^{-1}} } }   
\nonumber \\ 
&-q_0 \rho \psi
{\frac{6M-2\ell + 1+\rho \ell  }{ 2\prnt{1+\rho \ell}^2 \sqrt{3M^2-{\ell}{M}} \hspace{1mm} }}
\end{align}
By standard lines of derivation, one can show that for any choices of $\rho$, $\mar'(M) < 0$. This indicates that for all \ac{snr}s the eavesdropper is relatively prevailing. Moreover, for
\begin{align}
\rho > (1+a_\epsilon)\loge M + a_\epsilon -1 \label{eq:rho_M_out}
\end{align}
with $a_\epsilon \coloneqq -3q_0 \psi /\sqrt{2}$, the outage secrecy rate is a decreasing function of $L$, and therefore, $L^\star=1$. Nevertheless, when \eqref{eq:rho_M_out} does not hold, the optimal number of active transmit antennas antennas is given by\footnote{One should note that for $L^\star=1$, we have $\beta_\ee = 1$, and therefore, this approximation is not necessarily consistent. Nevertheless, as shown through numerical investigations, the approximation is accurate even in this regime.} $L^\star \approx \lfloor \ell^\star\rceil$ where $\ell^\star$ fulfills $\mar'(\ell^\star)=0$.
\begin{figure}[t]
\hspace*{-.75cm}  
%%\centering
\resizebox{1.1\linewidth}{!}{
\pstool[width=.35\linewidth]{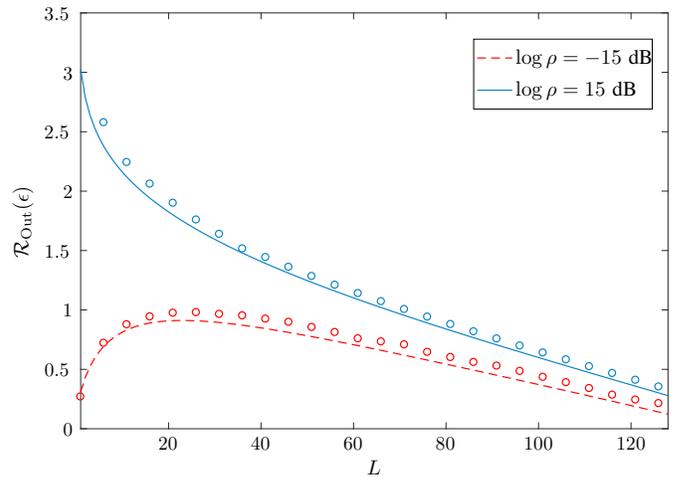}{
\psfrag{Outage}[c][t][0.27]{$\mar_\out(\epsilon)$}
\psfrag{Lt}[c][c][0.27]{$L$}
\psfrag{rho=-15AAA-SABA-AS}[l][l][0.27]{$\log \rho=-15$ dB}
\psfrag{rho=15AAA-SABA-AS}[l][l][0.27]{$\log \rho=15$ dB}

%y-axis
\psfrag{20}[c][b][0.25]{$20$}
\psfrag{40}[c][b][0.25]{$40$}
\psfrag{60}[c][b][0.25]{$60$}
\psfrag{80}[c][b][0.25]{$80$}
\psfrag{100}[c][b][0.25]{$100$}
\psfrag{120}[c][b][0.25]{$120$}

%x-axis
\psfrag{0}[r][c][0.25]{$0$}
\psfrag{0.5}[r][c][0.25]{$0.5$}
\psfrag{1}[r][c][0.25]{$1$}
\psfrag{1.5}[r][c][0.25]{$1.5$}
\psfrag{2}[r][c][0.25]{$2$}
\psfrag{2.5}[r][c][0.25]{$2.5$}
\psfrag{3}[r][c][0.25]{$3$}
\psfrag{3.5}[r][c][0.25]{$3.5$}
\psfrag{4}[r][c][0.25]{$4$}
\psfrag{4.5}[r][b][0.25]{$4.5$}
}}
\vspace{-8mm}\caption{The $\epsilon$-outage secrecy rate ($\epsilon=0.1$) versus $L$, for $M=128$. For the first case $\mar_\out (0.1)$ is a decreasing function of $L$, while for $\log \rho=-15$ dB both the simulation and approximation results suggest $L^\star=23$. The solid lines and the circles denote the approximation and simulation, respectively.
}
\label{fig:rrr}
%\vspace{-4.0mm}
\end{figure}
In Fig.~\ref{fig:rrr}, the $\epsilon$-outage secrecy rate at $\epsilon=0.1$ for $M=128$ has been plotted versus $L$ for $\log \rho=-15$ dB and~$\log \rho=15$~dB. As the figure depicts, for the latter case, in which the inequality in \eqref{eq:rho_M_out} is satisfied, $\mar_\out (\epsilon)$ is a decreasing function of $L$. For the case of $\log \rho=-15$ dB, the simulations indicate that $L^\star=23$. The analytic investigations moreover reports $\ell^\star=22.97$ which is consistent with the simulation results.

\end{example}

\section{Conclusion}
\label{conclusion}
%Conclusion
% has been investigated.
%de based on the eavesdropper's \ac{csi} availability at the transmitter. For active eavesdropping, in which the transmitter has knowledge about the \ac{csi} of the eavesdropper channel, 
% Also, for the passive eavesdropping scenario, we drive the asymptotic closed form expressions for the secrecy outage probability and $\epsilon$-outage secrecy which provide a good secrecy metric where the eavesdropper is purely passive and no \ac{csi} about eavesdropper is available at the transmitter. 
%However, if the legitimate receiver is relatively stronger than eavesdropper, the secrecy performance of the system in terms of ergodic secrecy rate and outage secrecy probability increase monotonically as the number of selected antennas increase. 

%For this setup, the instantaneous secrecy rate has been approximated in terms of a Gaussian random variable in the large-system limit. For the performance analysis, we have considered the both scenarios of active and passive eavesdropping and derived closed-form expressions for the achievable ergodic and the $\epsilon$-outage secrecy rate invoking our large-system results.
%We have then investigated the set of setup parameters for which antenna selection at the transmitter improves the secrecy performance in the both cases of active and passive eavesdropping. 
In this paper, we characterized the impacts of \ac{tas} on the secrecy performance of massive \ac{mimo} wiretap settings. It was shown that in some scenarios, the secrecy performance is enhanced under \ac{tas} compared to the case of full complexity. We moreover developed an analytic framework to determine the optimal number of active antennas in these~scenarios.~The numerical investigations confirmed the accuracy of our framework even for settings with not so large dimensions.~The~ana-lyses of this study implies that antenna selection in some mas-sive \ac{mimo} wiretap setups enhances the secrecy performance. A possible direction for future work is to extend the current framework to scenarios in which other techniques, such~as~art-ificial noise generation, are employed along with \ac{tas}~for~sec-recy enhancement. The work in this direction is ongoing. 
%concept of ``secrecy for free'' in the massive \ac{mimo} wiretap channels when a subset of transmit antennas are set to be active. Such investigations are currently underway.
%The current work can be further extended to investigate other possible impacts of antenna selection on physical layer security.

%than the legitimate receiver. For this case, using all the available antennas reduces the secrecy performance, and the secrecy performance metric is maximized for some number of active antennas less than the total number of available transmit antennas. 
%
%This discussion implies that for some setups, antenna selection enhances the secrecy performance of massive \ac{mimo} systems. The numerical investigations have shown that the optimal number of transmit antennas derived analytically from our large-system framework accurately matches the simulations even for not so large dimensions.

%therefore the result can be employed for analyzing secrecy performance measures on \ac{mimo} wiretap channels.

\begin{appendices}
\section{Derivation of Theorem \ref{thm:1}}
\label{app:1}
We start by evaluating the large-system distribution of $\mar_\mm$. It has been shown in \cite[Lemma~2]{asaad2017tas} that the distribution of the input-output mutual information of a Gaussian \ac{mimo} channel, under some constraints, is accurately approximated in terms of the random variables $\tr{\mJ}$ and $\tr{\mJ^2}$ where $\mJ\coloneqq\mH^\her \mH$. Under the \ac{tas} protocol $\mas$, $\tr{\mJ}$ represents the sum of $L$ first order statistics which at the large limit of $M$ converges in distribution to a Gaussian random variable whose mean and variance are given by \eqref{eq:eta_t} and \eqref{eq:sigma_t}, respectively. Using some properties of random matrices, the large-system distribution of $\mar_\mm$ is then approximated as in \cite[Theorem~1]{asaad2017tas} with a Gaussian distribution whose mean and variance are given in terms of $\eta_t$ and $\sigma_t^2$. The next step is to evaluate the distribution of $\mar_\ee$. Noting that the main and the eavesdropper channel are independent, it is concluded that the \ac{tas} protocol $\mas$ performs as a random selection protocol from the eavesdropper's point of view. By considering the asymmetrically asymptotic regime of eavesdropping, one can invoke the asymptotic results for \ac{iid} Gaussian fading channels in \cite{hochwald2004multiple}, and approximate the large-system distribution of $\mar_\ee$ is with a Gaussian distribution whose mean and variance respectively~read
\begin{subequations}
\begin{align}
\hspace*{-2mm} \eta_\ee &= U_\ee \log\prnt{1+\rho_\ee V_\ee}\\
\hspace*{-2mm}\sigma_\ee^2 &= \left( \mone_{\set{N_\ee > L}}\frac{U_\ee}{V_\ee}\hspace*{-.7mm} +\hspace*{-.7mm}\mone_{\set{N_\ee < L}} \frac{U_\ee V_\ee \rho^2_\ee}{\left(1+\rho_\ee V_\ee\right)^2  } \right) \psi^2.
\end{align}
\end{subequations}
Since the main and the eavesdropper channel are independent, $\mar^\star=\mar_\mm-\mar_\ee$ is sum of two independent Gaussian random variables in the large-system limit; hence, it is Gaussian with mean and variance given in \eqref{eq:eta_final} and \eqref{eq:sigma_final}.
%With \cite[Lemma~1]{asaad2017tas} and taking the limit of $M\uparrow\infty$, Therefore, $\tr{\mJ}$ and $\tr{\mJ^2}$, in the asymmetrically asymptotic regime of eavesdropping, can be determined explicitly as functions of independent Gaussian random variables.
\section{Derivation of Theorem \ref{thm:2}}
\label{app:2}
In the large-system limit, the achievable ergodic and~$\epsilon$-out-age secrecy rate are accurately approximated by \eqref{eq:asy} and \eqref{eq:R_out-Final}, respectively. To derive a sufficient condition, we first consider Scenario~A. We define the function $\mam_{\rm A}(\cdot)$ on the real~axis~as
\begin{align}
\mam_{\rm A}(\ell) = s(\ell) \hspace*{.5mm} \phi\prnt{\frac{f(\ell)}{s(\ell)}}+ f(\ell) \hspace*{.5mm}\rmQ\prnt{-\frac{f(\ell)}{s(\ell)}}
\end{align}
where $f(\ell)$ and $s(\ell)$ are determined by replacing $L$ with $\ell$ in the asymptotic terms given for $\eta$ and $\sigma$ in Theorem~\ref{thm:1}, respectively. $\mam_{\rm A}(\ell)$ is the real envelope of the achievable ergodic rate whose values at integer points within the interval $[1,M]$ give $\mar_\rmE(\mas)$. It is therefore concluded that for the set of $T=(\rho_\mm,\rho_\ee, N_\rr, N_\ee)$ in which $\mam_{\rm A}(\ell)$ is an increasing function, the legitimate receiver is relatively prevailing. In this case, a set of sufficient conditions are deduced by investigating the set of tuples in which $\mam_{\rm A}'(\ell)>0$ for all $\ell\in [1,M]$. To extend the result to Scenario B, one may similarly define %the function~$\mam_{\rm P}(\ell)$~as 
\begin{align}
\mam_{\rm P}(\ell) = f(\ell) - \rmQ^{-1} \prnt{\epsilon} s(\ell) 
\end{align}
and investigate a sufficient condition for which $\mam_{\rm P}'(\ell)>0$. 

\begin{lemma}
\label{lemm:1}
Assume that $f'(\ell)>0$ for $\ell\in [1,M]$. Then, %we have
\begin{itemize}
\item[(a)] $s'(\ell)\approx 0$ for $\ell< \min\set{ N_\ee , N_\rr}$, and
\item[(b)] $f(\ell)/s(\ell) \gg 1$ for $\ell\in [1,M]$.
\end{itemize}
\end{lemma}

\begin{prf}
Let $f'(\ell)>0$. In this case, for $\ell< \min\set{ N_\ee , N_\rr}$, one can simply show that
\begin{align}
N_\ee {\rho_\ee} <  \frac{N_\rr \rho_\mm u}{N_\rr +\rho_\mm \eta_t} \prnt{1+\rho_\ee \ell} \label{eq:Constraint}
\end{align}
where $u$ and $\eta_t$ are defined in Theorem~\ref{thm:2}. As we have assumed an asymmetrically asymptotic regime of eavesdropping, the number of eavesdropper antennas reads\footnote{Definition~\ref{def:rel} indicates that in this regime $\beta_\ee=N_\ee/L$ reads either $\beta_\ee \ll 1$ or $\beta_\ee \gg 1$.} $N_\ee \gg \ell$ when $\ell< \min\set{ N_\ee , N_\rr}$. This means that the inequality in \eqref{eq:Constraint} holds only when $u$ takes values close to zero. By taking the first derivative of $s(\ell)$, it is then shown that for values of $u$ close to zero, $s'(\ell)\approx 0$ for  $\ell< \min\set{ N_\ee , N_\rr}$ which concludes (a). 

To prove (b), one may note that for $\ell> \min\set{ N_\ee , N_\rr}$ we have $s'(\ell) < 0$. This statement along with (a) depicts that 
\begin{align}
\frac{\partial}{\partial \ell} \left( \frac{f(\ell)}{s(\ell)} \right) >  0
\end{align}
when $f'(\ell) >0$. As $f(1)/s(1) \gg 1$ for large $M$, one~concludes that $f(\ell)/s(\ell) \gg 1$ for $\ell\in [1,M]$.% which concludes (b).
\end{prf}

From Lemma~\ref{lemm:1}, it is observed that $f'(\ell)>0$ is a sufficient condition in the asymptotic regime to have both $\mam_{\rm A}'(\ell)>0$ and $\mam_{\rm P}'(\ell)>0$. In fact by using Part (b) in Lemma~\ref{lemm:1}, we employ $\rmQ(-\xi)\approx 1-{ \phi(\xi)}/{\xi}$ and write $\mam_{\rm A}(\ell) = f(\ell)$ when $f'(\ell)>0$. This concludes the proof for Scenario~A. Moreover, %one can write 
\begin{align}
\mam_{\rm P}'(\ell) = h'(\ell) - \rmQ^{-1} \prnt{\epsilon} s'(\ell). \label{eq:akhar}
\end{align}
%By chain derivative,
%\begin{align}
%\mam_{\rm A}(\ell) = s'(\ell) \hspace*{.5mm} \phi\prnt{\frac{f(\ell)}{s(\ell)}}+ h'(\ell) \hspace*{.5mm}\rmQ\prnt{-\frac{f(\ell)}{s(\ell)}}.
%\end{align}
Note that $s'(\ell) < 0$ for $\ell\in [\min\set{ N_\ee , N_\rr},M]$. Hence, \eqref{eq:akhar} along with Part (a) in Lemma~\ref{lemm:1} implies that $f'(\ell)>0$ is a sufficient condition for Scenario B as well. Finally, by taking the derivative of $f'(\ell)$ and noting that $\partial \eta_t / \partial \ell =u$, the proof is completed.
\end{appendices}

%In order to find the exact asymptotic characteristics of the random variable $\tr{\mJ}$ in Section \ref{sec:main}, we invoke the result reported in \cite{stigler1973asymptotic}. Using the main theorem of \cite{stigler1973asymptotic}, $\eta_t$ reads
%\begin{align}
%\eta_t=N_\mathrm{r}\left[ L  + M f_{N_\mathrm{r} +1}(u) \right]
%\end{align}
%where $f_{N_\mathrm{r}}(\cdot)$ denotes the chi-square probability density function with $2N_\mathrm{r}$ degrees of freedom and mean $N_\mathrm{r}$,
% \begin{equation}
% \label{eq:Je}
%    f_{N_\mathrm{r}}(x)= \frac{1}{(N_\mathrm{r}-1)!}
%    \begin{cases}
%     x^{N_\mathrm{r}-1} e^{-x} , & \text{if}\ x \geq 0 \\
%      0, & \text{if}\ x < 0
%    \end{cases}
%  \end{equation}
%and $u$ is the solution of the equation
%\begin{align}
%\int_u^\infty f_{N_\mathrm{r}}(x) \mathrm{d} x= \frac{L}{M}.
%\end{align}
%Moreover, $\sigma_t^2$ is determined as
%\begin{align}
%\sigma_t^2=\left(Lu-\eta_t \right)^2 \left(\frac{1}{L}-\frac{1}{M} \right) - \frac{\eta_t^2}{L}+ \Xi_t
%\end{align}
%where the non-negative scalar $\Xi_t$ is defined as
%\begin{align}
%\Xi_t= N_\mathrm{r} \left(N_\mathrm{r}+1\right) \left[ L+M f_{N_\mathrm{r}+1}(u)+M f_{N_\mathrm{r}+2}(u) \right].
%\end{align}

\bibliographystyle{IEEEtran}
\bibliography{ref}

\end{document}

%% file: commands.tex
\newcommand{\setS}{\mathbbmss{S}}
\newcommand{\setW}{\mathbbmss{W}}

\newcommand{\setC}{\mathbbmss{C}}

\newcommand{\tmH}{\tilde{\mathbf{H}}}

\newcommand{\argmax}{\mathop{\mathrm{argmax}}}

\newcommand{\rec}{\mathrm{R}}
\newcommand{\eve}{\mathrm{E}}

\newcommand{\out}{\mathrm{Out}}

\newcommand{\rmQ}{\mathrm{Q}}

\newcommand{\her}{\mathsf{H}}

\newcommand{\bfh}{\mathbf{h}}

\newcommand{\rr}{\mathrm{r}}

\newcommand{\ee}{\mathrm{e}}
\newcommand{\mm}{\mathrm{m}}

\newcommand{\mas}{\mathcal{S}}

\newcommand{\mam}{\mathcal{M}}

\newcommand{\mar}{\mathcal{R}}
\newcommand{\Pout}{\mathcal{P}}

\newcommand{\bx}{{\boldsymbol{x}}}

\newcommand{\set}[1]{\left\lbrace#1\right\rbrace}
\newcommand{\cnormal}[1]{\mathcal{CN}\left( \mathbf{0}, #1 \right)}

\newcommand{\bz}{{\boldsymbol{z}}}

\newcommand{\by}{{\boldsymbol{y}}}

\newcommand{\bn}{{\boldsymbol{n}}}

\newcommand{\mI}{\mathbf{I}}
\newcommand{\mone}{\mathbf{1}}
\newcommand{\mJ}{\mathbf{J}}

\newcommand{\mQ}{\mathbf{Q}}

\newcommand{\mH}{\mathbf{H}}

\newcommand{\rms}{\mathrm{s}}
\newcommand{\rmE}{\mathrm{Erg}}
\newcommand{\asy}{\mathrm{asy}}
\newcommand{\rmO}{\mathrm{o}}
\newcommand{\nzs}{\mathrm{NZS}}
\newcommand{\E}{\mathbb{E}\hspace{.2mm}}

\newcommand{\norm}[1]{\lVert #1 \rVert}
\newcommand{\prnt}[1]{\left( #1 \right)}

\newcommand{\abs}[1]{\lvert #1 \rvert}
\newcommand{\tr}[1]{\mathrm{Tr} \{ #1 \}}
\newcommand{\loge}[1]{\mathrm{ln}\hspace*{.5mm} #1 }

\newtheoremstyle{mystyle}%                % Name
  {}%                                     % Space above
  {}%                                     % Space below
  {}%                                     % Body font
  {}%                                     % Indent amount
  {\bfseries}%                            % Theorem head font
  {:}%                                     % Punctuation after theorem head
  { }%                                    % Space after theorem head, ' ', or \newline
  {}%                                     % Theorem head spec (can be left empty, meaning `normal')

\theoremstyle{mystyle}

\newtheorem{definition}{Definition}

\newtheorem{theorem}{Theorem}
\newtheorem*{remark}{Remark}
\newtheorem*{prf}{Proof}
\newtheorem{example}{Example}
\newtheorem{lemma}{Lemma}
\newcounter{bar}